\newcommand{\avg}[1]{\left< #1 \right>} 

\documentclass[letterpaper]{aastex}
\usepackage{amsbsy}
\usepackage{amsfonts}
\usepackage{amssymb}
\usepackage{lscape}
\usepackage{graphicx}
\usepackage{graphics}
\usepackage{dcolumn}
\usepackage{bm}
\begin{document}


\title{Evidence for Wave Heating of the Quiet Sun Corona}
\author{M. Hahn\altaffilmark{1} and D. W. Savin\altaffilmark{1}}

\altaffiltext{1}{Columbia Astrophysics Laboratory, Columbia University, MC 5247, 550 West 120th Street, New York, NY 10027 USA}

\date{\today}
\begin{abstract}
	
	We have measured the energy and dissipation of Alfv\'enic waves in the quiet Sun. A magnetic field was used to infer the location and orientation of the magnetic field lines along which the waves are expected to travel. The waves were measured using spectral lines to infer the wave amplitude. The waves cause a non-thermal broadening of the spectral lines, which can be expressed as a non-thermal velocity $v_{\mathrm{nt}}$. By combining the spectroscopic measurements with this magnetic field model we were able to trace the variation of $v_{\mathrm{nt}}$ along the magnetic field. At the footpoints of the quiet Sun loops we find that waves inject an energy flux in the range of $1.2$--$5.2 \times 10^{5}$~$\mathrm{erg\,cm^{-2}\,s^{-1}}$. At the minimum of this range, this amounts to more than 80\% of the energy needed to heat the quiet Sun. We also find that these waves are dissipated over a region centered on the top of the loops. The position along the loop where the damping begins is strongly correlated with the length of the loop, implying that the damping mechanism depends on the global loop properties rather than on local collisional dissipation. 

\end{abstract}

\maketitle
	
\section{Introduction}\label{sec:intro}

	One of the major theories to explain coronal heating, is that the heating is caused by the dissipation of magnetohydrodynamic waves that are launched into the corona by agitation in and below the photosphere \citep{Cranmer:SSR:2002}. The most likely waves to carry this energy are Alfv\'enic waves, which have been detected throughout the solar atmosphere \citep{Belcher:JGR:1971,Tomczyk:Sci:2007,McIntosh:Nature:2011}. In order to ascertain if these waves actually do heat the corona, observations are needed to determine both whether the waves carry enough energy into the corona and if they lose sufficient energy at the appropriate heights. 
		
	There are several methods that have been used to detect waves via spectroscopy. One method is to observe spectra at a high cadence so that the Doppler shifts induced by the wave motions can be resolved. Such measurements have been carried out with the Coronal Multichannel Polarimeter \citep[CoMP;][]{Tomczyk:SolPhys:2008}. These data show that Alfv\'enic waves are present in quiet Sun regions. By studying the energy flow along quiet Sun coronal loops, CoMP measurements have found that more energy is present in the upward propagating waves than in the downward propagating waves. Since waves are expected to be launched from the footpoints, this implies that the waves are dissipated near the top of the loop. More recent measurements have shown that the power spectrum of waves near the top of the quiet Sun loops is broader than at the base of the loops, which suggests that there is turbulence at the loop top that may cause wave damping. However, a limitation for these measurements is that the energy content of the waves is believed to be underestimated due to line of sight effects \citep{McIntosh:ApJ:2012}. The integration along the line of sight tends to wash out the observed Doppler shifts, which are therefore much smaller than the actual wave amplitudes leading to an underestimate of the total wave power. 
	
	Non-thermal broadening of spectral lines is another diagnostic for waves in the corona. This broadening is caused by the bulk fluid motions induced by the waves and is proportional to the wave amplitude. Thus, the wave energy flux can be determined from the inferred amplitudes. Moreover, spatial variations of the amplitudes can show if the waves are damped and indicate the dissipation rate. This type of diagnostic was the basis for several coronal hole studies demonstrating that waves heat the corona and accelerate the fast solar wind in coronal holes \citep{Bemporad:ApJ:2012, Hahn:ApJ:2012, Hahn:ApJ:2013a}. 
		
	There are also several existing line width measurements for quiet Sun regions. Some of these observations have suggested that lines become narrower with increasing height, which is consistent with a decreasing amplitude of the waves and therefore wave dissipation \citep{Hassler:ApJ:1990, Harrison:AA:2002}. But, other observations have not found such line narrowing \citep{OShea:AA:2003, Wilhelm:AA:2004}. A possible explanation for these discrepancies is that the existing measurements have not considered the geometry of the magnetic field, but simply observed line widths as a function of height. Since Alfv\'en waves in the corona are expected to propagate along the magnetic field lines, the apparent damping may depend on the coordinate along the quiet Sun loops, rather than height.
	
	Here, we combine spectroscopic measurements with a magnetic field model in order to study the variation of line widths along the magnetic fields. Section~\ref{sec:obs} presents the observations and Section~\ref{sec:ana} describes our analysis techniques. The results concerning the energy content of the waves and their dissipation are discussed in Section~\ref{sec:res}. A summary is given in Section~\ref{sec:con}.
	
\section{Observation} \label{sec:obs}

\begin{figure}
\centering
\includegraphics[width=\textwidth]{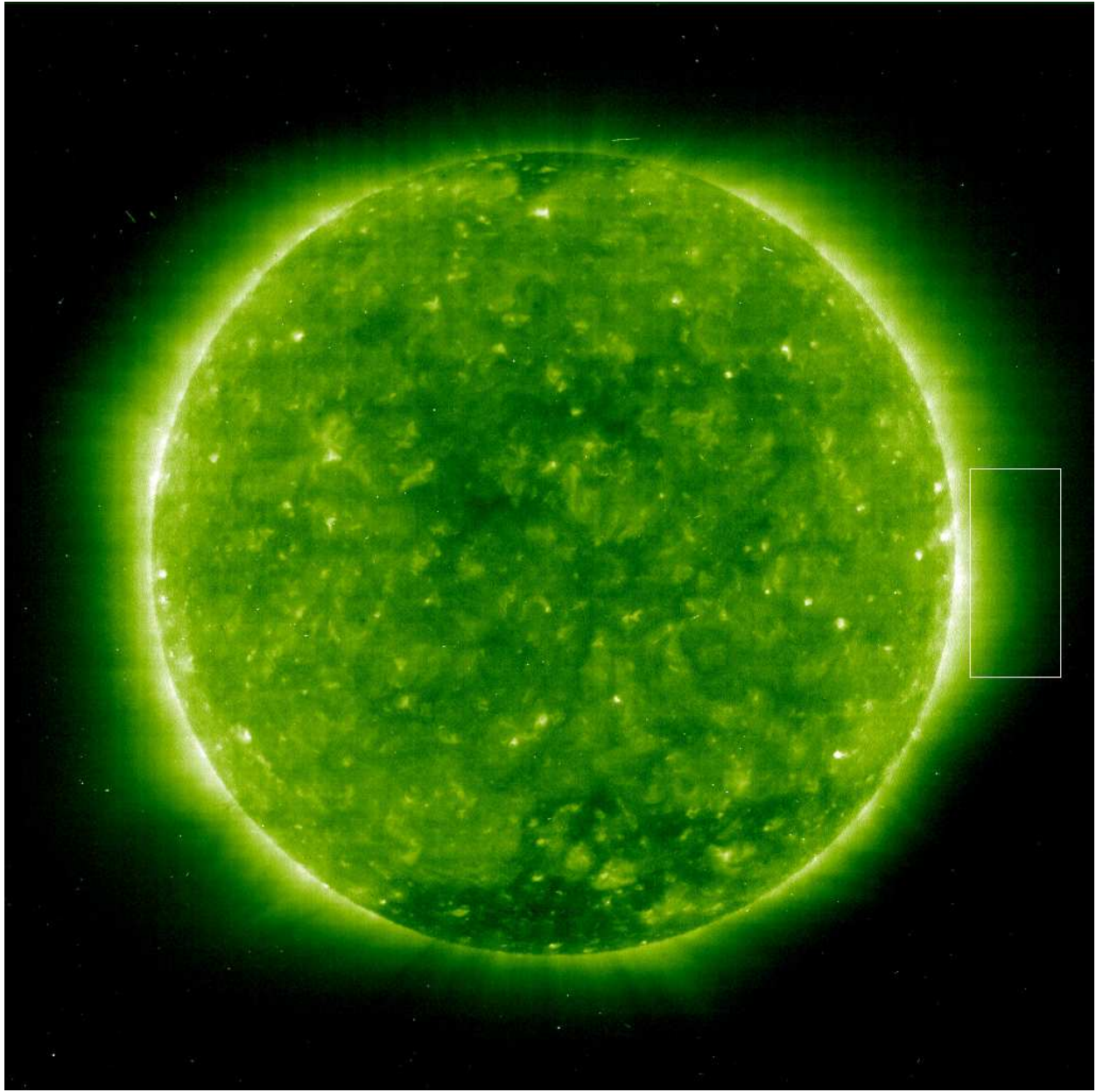}
\caption{\label{fig:context1} The box outlines the position of the EIS observation made on 21-Jan-2008 overlayed on an EIT \textit{SOHO} image in the 195~\AA\ band.
}
\end{figure}

\begin{figure}
\centering 
\includegraphics[width=\textwidth]{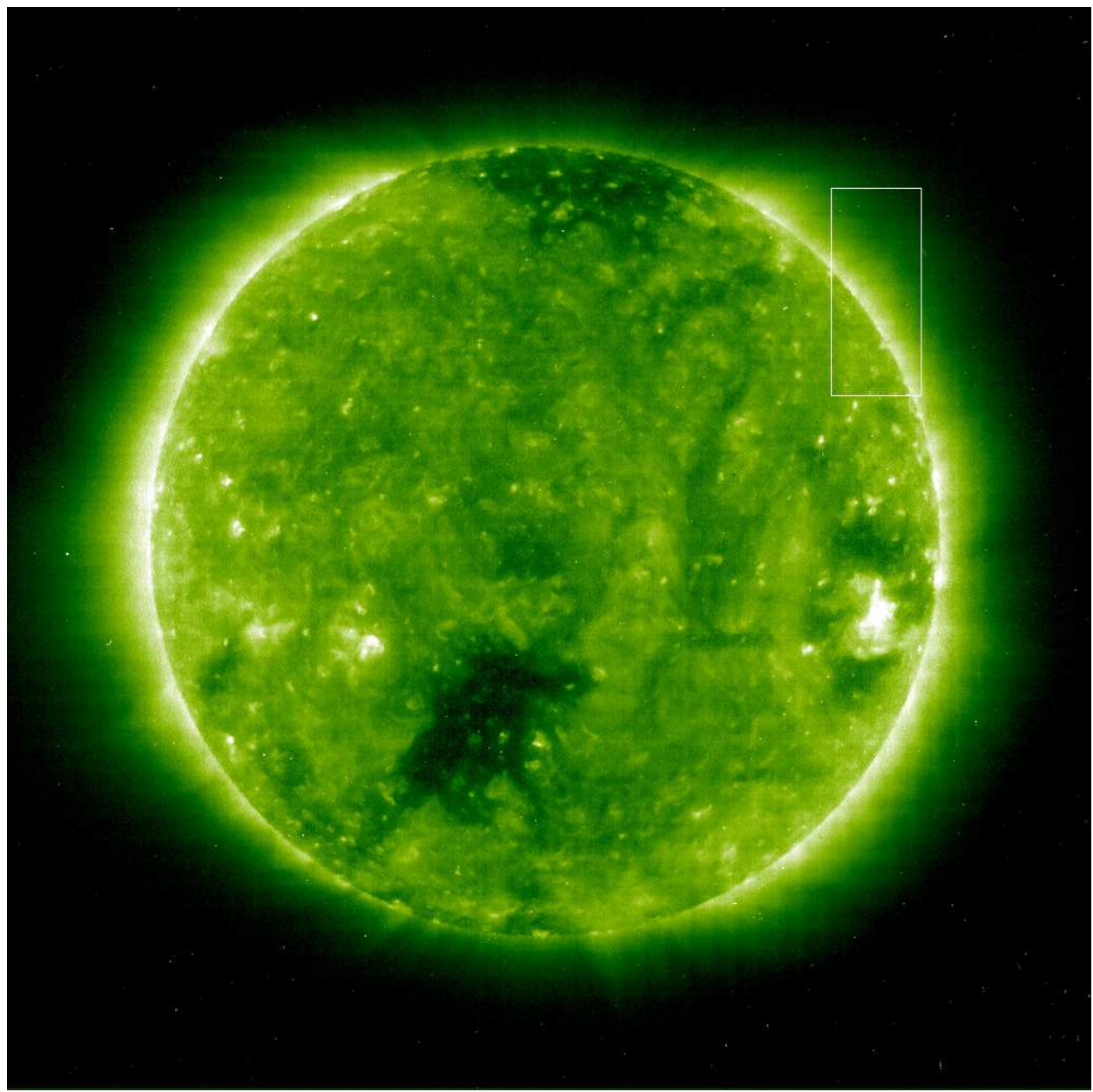}
\caption{\label{fig:context2} Same as Figure~\ref{fig:context1}, but for the EIS observation on 21-Oct-2007.
}
\end{figure}

	We analyzed spectra from two observations made with the Extreme Ultraviolet Imaging Spectrometer \citep[EIS;][]{Culhane:SolPhys:2007} on \textit{Hinode}. One of these observations focused on an equatorial region of the corona and the other on a higher latitude region. The equatorial observation was made on 2008 January 21 at 11:45 UT, while the higher latitude region data were taken on 2007 October 21 at 17:11 UT. Figures~\ref{fig:context1} and \ref{fig:context2} show the location of these observations superimposed on images taken at about the same times using the Extreme ultraviolet Imaging Telescope \citep[EIT;][]{Delaboudiniere:SolPhys:1995}. 

	In both cases, the EIS observations used the 1$^{\prime\prime}$ slit and covered the full EIS wavelength range. Both observations were rastered across 22 positions with a raster step size of about $10^{\prime\prime}$, thus covering a width in the X-direction of about $220^{\prime\prime}$. The exposure time at each raster position was 60~s. 
	
	The data were prepared using the \textit{eis\_prep} procedure to remove spikes, warm pixels, and CCD dark current, and calibrate the intensity scale. Drifts in the wavelength scale were corrected using the method of \citet{Kamio:SolPhys:2010}. After making these corrections, the data were binned over eight pixels in the vertical direction in order to increase the signal-to-noise ratio. In the following discussion, we refer exclusively to these binned data. 

\section{Analysis} \label{sec:ana}
\subsection{Intensities and Line Widths} \label{subsec:fit}

	In order to extract the line intensity, line width $\Delta \lambda$, and line centroid, we fit a Gaussian profile to each spectral line. A broad selection of lines was used to characterize the density and temperature structure in the observation. For these aspects of the analysis, the line list was essentially the same as that used in \citet{Hahn:ApJ:2012}. 
	
	Line widths are more sensitive to noise than intensities, so for line width measurements we selected six bright, unblended lines from the spectrum. These lines were observed to heights of 1.25~$R_{\sun}$. The selected lines were Si~\textsc{x} 271.99~\AA, Fe~\textsc{ix} 197.86~\AA, Fe~\textsc{x} 184.54~\AA, Fe~\textsc{xi} 188.22~\AA, Fe~\textsc{xii} 195.12~\AA, and Fe~\textsc{xiii} 202.04~\AA. For the Fe~\textsc{xii} line, there is another line from the same ion at about 195.18~\AA\ that is far enough away in wavelength to be seperated using a double-Gaussian fit.

	The line width depends on instrumental broadening $\Delta \lambda_{\mathrm{inst}}$, ion temperature $T_{\mathrm{i}}$ and the non-thermal velocity $v_{\mathrm{nt}}$. The full width at half maximum (FWHM) of the line $\Delta \lambda_{\mathrm{FWHM}}$ is given by \citep{Phillips:book} 
\begin{equation}
\Delta \lambda_{\mathrm{FWHM}} = \left[ \Delta\lambda_{\mathrm{inst}}^2 +  
4 \ln(2)\left(\frac{\lambda}{c}\right)^{2}\left(\frac{2k_{\mathrm{B}}T_{\mathrm{i}}}{M} + v_{\mathrm{nt}}^2 \right) \right]^{1/2}, 
\label{eq:width}
\end{equation}	
where $M$ is the ion mass, $k_{\mathrm{B}}$ is the Boltzmann constant, $\lambda$ is the observed wavelength, and $c$ is the speed of light. The instrumental width varies with position along the slit. We have subtracted the instrumental width using the new calibration given by \citet{Hara:2014}, which was obtained by cross calibrating EIS line widths with measurements from ground-based spectra. The remaining line width is then the sum of the thermal and non-thermal velocities and can be expressed in terms of an effective velocity, defined as
\begin{equation}
v_{\mathrm{eff}}=\sqrt{\left(v_{\mathrm{th}}^2 + v_{\mathrm{nt}}^2 \right)},
\label{eq:veffdefine}
\end{equation}
where the thermal velocity is $v_{\mathrm{th}}= 2k_{\mathrm{B}}T_{\mathrm{i}}/M$. 

\subsection{Electron Density and Temperature} \label{subsec:dem}
	
	The density was measured using line intensity ratios. Two diagnostics were used, Fe~\textsc{xii} 195.12~\AA/196.64~\AA\ and Fe~\textsc{xiii} 202.04~\AA/(203.79~\AA\ $+$ 203.82~\AA). For both observations, the Fe~\textsc{xii} measurements show that the density decreases from about $\sim 5 \times 10^{8}$~$\mathrm{cm^{-3}}$ at 1.05~$R_{\sun}$ to $\lesssim 1 \times 10^{8}$~$\mathrm{cm^{-3}}$ at $1.25$~$R_{\sun}$. The Fe~\textsc{xiii} density diagnostic infers a spatial dependence similar to the Fe~\textsc{xii} diagnostic, but the magnitude of the density is smaller by about a factor of two. Such discrepancies between these two density diagnostics are well known, and have been attributed to uncertainties in the underlying atomic data \citep{Young:AA:2009}. For the line width analysis, we have used the Fe~\textsc{xii} results, but we have also estimated the systematic uncertainties in our results due to this discrepancy in the density measurements. 
	
	The temperature structure of the observations was characterized using a differential emission measure (DEM) analysis. The DEM $\phi(T_{\mathrm{e}})$ describes the amount of material along the line of sight at the temperature $T_{\mathrm{e}}$. In terms of the DEM, the intensity of an emission line emitted by a transition from level $j$ to level $i$ is given by 
\begin{equation}
I_{ji} = \frac{1}{4\pi}\int{ G(T_{\mathrm{e}})\phi(T_{\mathrm{e}}) dT_{\mathrm{e}}},
\label{eq:intensity}
\end{equation}
where $G(T_{\mathrm{e}})$ is the contribution function and describes the level populations, ionization balance, elemental abundance, and radiative decay rates \citep{Phillips:book}. We used the CHIANTI atomic database to obtain these parameters \citep{Dere:AA:1997, Landi:ApJ:2013}. Given $G(T_{\mathrm{e}})$ and a set of measured line intensities $I_{ji}$, it is possible to invert Equation~(\ref{eq:intensity}) and derive $\phi(T_{\mathrm{e}})$. 

	A variety of methods have been developed for performing the inversion. For example, \citet{Landi:AA:1997} have presented an iterative method, while \citet{Hannah:AA:2012} have published a code that uses regularized inversion technique. Another method is to model $\phi(T_{\mathrm{e}})$ using a function having several free parameters. These parameters are then adjusted using a least squares algorithm until the intensities calculated using Equation~(\ref{eq:intensity}) best match the observed intensities. 
	
	We performed the DEM analysis for the various spatial bins in our data using several methods. In most cases the DEM consisted of a single strong peak. That is, most of the region observed is roughly isothermal with some spread in temperature around the peak. The exception to this isothermality, is that there are some areas where there are prominences. In those locations, the DEM has a low temperature tail. These prominences are clearly visible in the He~\textsc{ii}~304~\AA\ images taken with EIT. As explained below in Section~\ref{subsec:trace}, these regions are eventually omitted from the analysis.
	
	Since the detailed DEM analysis shows that most of the observation can be described by a peak with some spread around it, we have adopted a simplified function for the DEM in order to more easily characterize the temperature structure. We assume that $\phi(T_{\mathrm{e}})$ can be represented as a Gaussian-like function: 
\begin{equation}
\phi(T_{\mathrm{e}}) = \frac{\mathrm{EM}}{\sigma_{T} \sqrt{2\pi}} \frac{1}{T_{\mathrm{e}}} \exp\left\{\frac{-[\log(T_{\mathrm{e}})-\mu_{T}]^2}{2\sigma_{T}^2}\right\}.  
\label{eq:demfunct}
\end{equation}
The DEM inversion was performed by using a least-squares fit to the measured intensities to find the free parameters EM, $\mu_{\mathrm{T}}$, and $\sigma_{\mathrm{T}}$. Equation~(\ref{eq:demfunct}) is the function for a log-normal distribution, and the fit parameters have a clear physical interpretation. EM is the total emission measure, i.e., $\phi(T_{\mathrm{e}})$ integrated over temperature. The peak of $\phi(T_{\mathrm{e}})$ occurs at the temperature $\log T_{\mathrm{e}} = \mu_{T} - \sigma_{T}^2$ and the spread of the distribution is characterized by $\sigma_{T}$. 
	
\begin{figure}
\centering \includegraphics[width=0.9\textwidth]{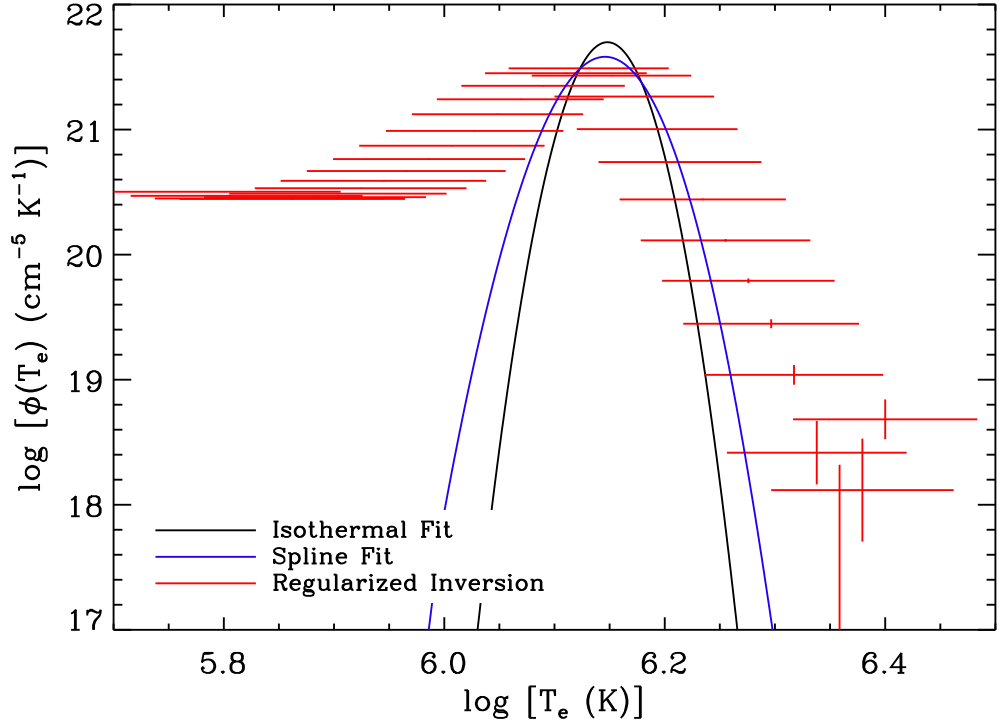}
\caption{\label{fig:isothermal} DEM distribution in the equatorial observation at the position (1039$^{\prime\prime}$, -238$^{\prime\prime}$) relative to Sun center. The black curve corresponds to the fit to Equation~(\ref{eq:demfunct}), the blue curve is the DEM derived from a more general spline function, and the red crosses come from a regularized inversion method. This DEM indicates that the plasma along the line of sight at this position is largely isothermal. The low temperature plateau in the spline fit is a systematic error due to the lack of constraints on the DEM at low temperatures. 
}
\end{figure}

\begin{figure}
\centering \includegraphics[width=0.9\textwidth]{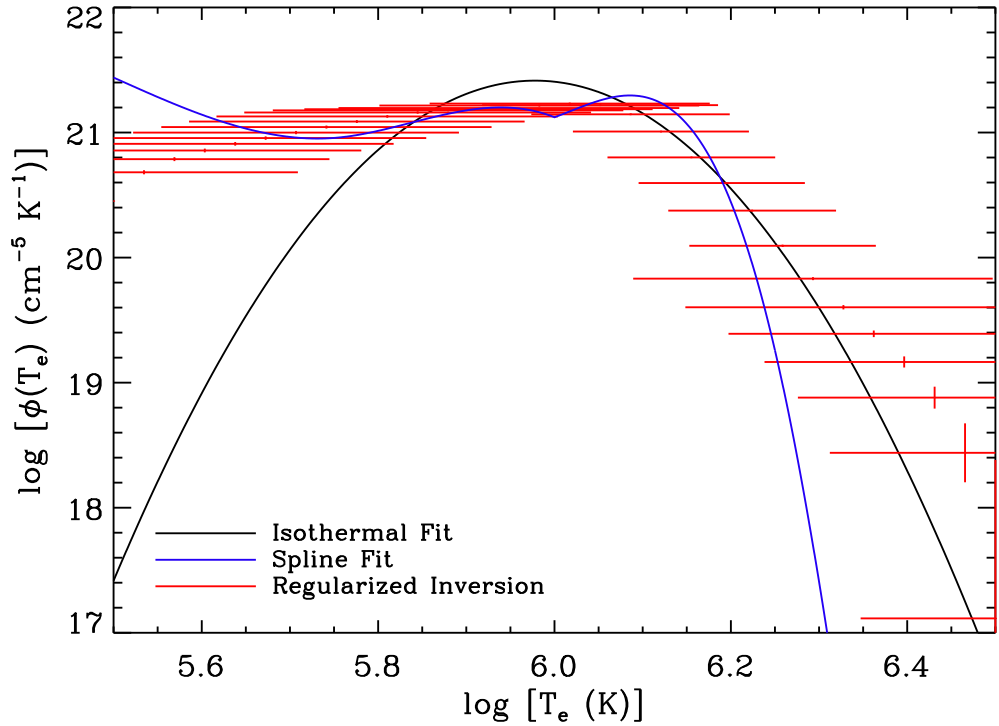}
\caption{\label{fig:multithermal} Same as Figure~\ref{fig:isothermal}, but for the position (1039$^{\prime\prime}$, 106$^{\prime\prime}$). Here the DEM implies that the emission is multithermal, coming from a broad distribution of temperatures along the line of sight. 
}
\end{figure}

\begin{figure}
\centering \includegraphics[width=0.9\textwidth]{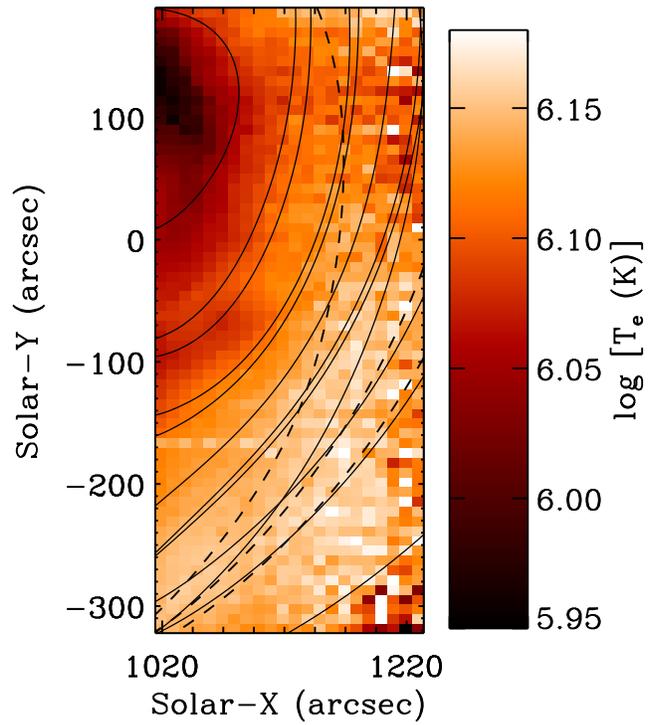}
\caption{\label{fig:region41} DEM peak temperature map with selected field lines overlayed for the 21-Jan-2008 equatorial observation. The dashed curves trace the three field lines used as examples later in Figure~\ref{fig:fafit}. This map shows the results for the entire field of view, without omitting the multithermal areas. 
}
\end{figure}

\begin{figure}
\centering \includegraphics[width=0.9\textwidth]{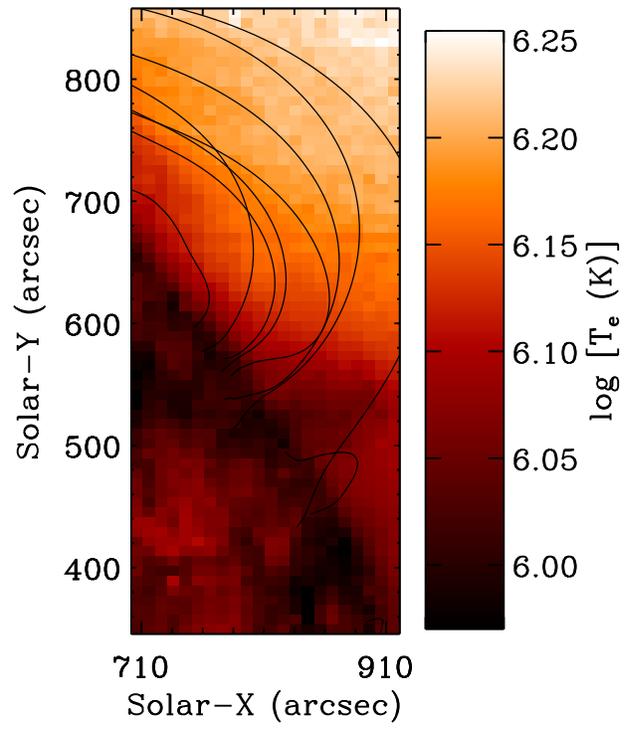}
\caption{\label{fig:region47} Same as Figure~\ref{fig:region41}, but for the 21-Oct-2007 high latitude observation. 
}
\end{figure}

	Figures~\ref{fig:isothermal} and \ref{fig:multithermal} illustrate the DEM analysis for two locations in the equatorial observation. Several different DEM inversion methods have been used. The black curves in the figures show the DEM derived from a fit to Equation~\ref{eq:demfunct}. The blue curves show a more general DEM that was found by performing a least squares fit to a spline function where the $\phi$ values of the spline knots served as the fit parameters. The red crosses indicate the results of the regularized inversion method of \citet{Hannah:AA:2012}. The DEM for Figure~\ref{fig:isothermal} indicates that the plasma is isothermal with a narrow peak at $\log T_{\mathrm{e}} = 6.15 \pm 0.03$. (Here and throughout all temperatures are given in units of Kelvin and uncertainties are given at an estimated $1\sigma$ statistical confidence level.) The spline fit seems to suggest a contribution at very low temperatures around $\log T_{\mathrm{e}} =5.5$, but this is probably not real and is instead an indication that the DEM inversion is not well constrained by the data at low temperatures. Such issues with DEM inversions have been discussed in detail by, e.g., \citet{Winebarger:ApJ:2012}. The DEM for Figure~\ref{fig:multithermal} implies that the plasma is multithermal, with a broad distribution of material at temperatures from $\log T_{\mathrm{e}} \approx 5.6$--$6.2$. Such multithermal DEMs are an indicator that there are multiple emitting structures along the line of sight, as discusssed in more detail in Section~\ref{subsec:trace}. Figures~\ref{fig:region41} and \ref{fig:region47} show maps of the peak temperature throughout the equatorial and high latitude observations, respectively, where we are still including multithermal areas at this point. 
	
\subsection{Ion Temperature}\label{subsec:ti}

\begin{figure}
\includegraphics[width=0.9\textwidth]{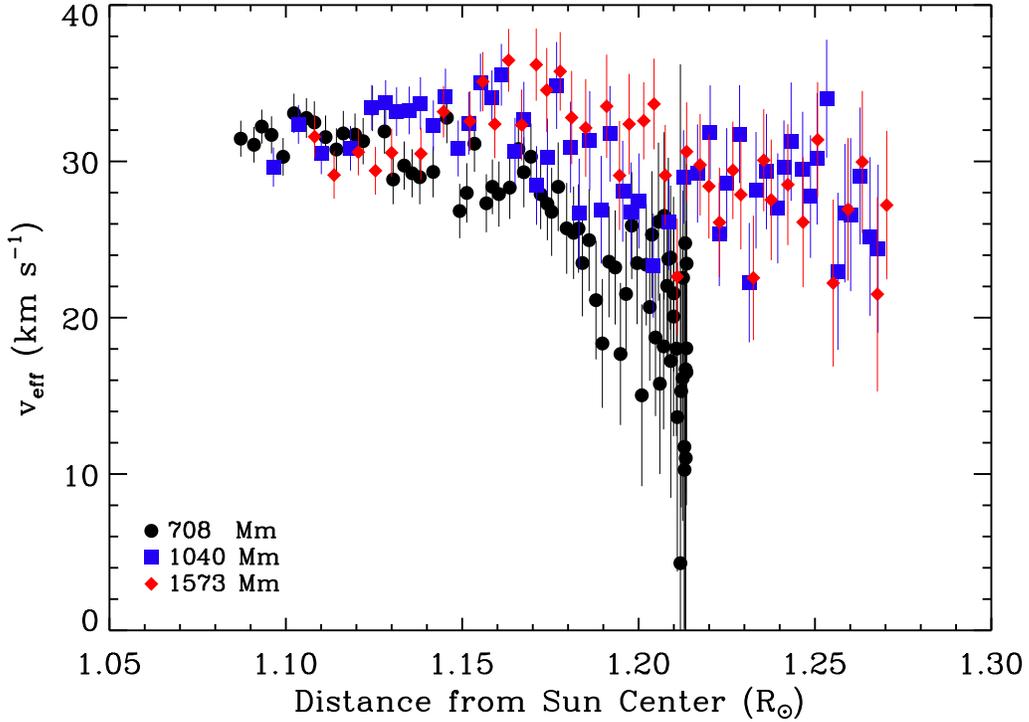}
\caption{\label{fig:veffvsr} The quantity $v_{\mathrm{eff}}$ for the Fe~\textsc{xii} line traced along three different magnetic field lines versus the height at each position along the field line. The three field lines are labelled according to their half-length $L$. They are the same as the field lines highlighted by the dashed lines in Figure~\ref{fig:region41} and shown in the more detailed analysis in Figure~\ref{fig:fafit}. The decrease of $v_{\mathrm{eff}}$ with increasing height implies wave damping.
}
\end{figure}

	Figure~\ref{fig:veffvsr} illustrates the measured spectral line width $v_{\mathrm{eff}}$ derived from the Fe~\textsc{xii} line as a function of height. The points plotted have been selected by tracing along three magnetic field lines, as described below. The three field lines for which the data are traced are labelled according to their half-length $L$. These field lines are the same ones that are highlighted by the dashed lines in Figure~\ref{fig:region41}. The plot shows that $v_{\mathrm{eff}}$ increases initially, and then decreases as the height increases. This implies that the wave amplitudes diminish with height and the waves are damped \citep{Bemporad:ApJ:2012, Hahn:ApJ:2012, Hahn:ApJ:2013a}. In order to measure this damping more quantitatively it is necessary to determine $v_{\mathrm{nt}}$. 

	The total line width depends on both $v_{\mathrm{nt}}$ and $T_{\mathrm{i}}$. Thus, it is necessary to determine $T_{\mathrm{i}}$ in order to study $v_{\mathrm{nt}}$. One approach is to find upper and lower bounds on $T_{\mathrm{i}}$ \citep{Tu:ApJ:1998}. This requires a few assumptions. The first assumption is that all of the ions have the same $v_{\mathrm{nt}}$. The basis for this assumption is that $v_{\mathrm{nt}}$ is caused by fluid motions, which affect all the ions in the same way, and that all of the emission comes from the same volume. Then an upper bound for $T_{\mathrm{i}}$ can be found by assuming that $v_{\mathrm{nt}}=0$. Given a set of line width measurements, a lower bound for $T_{\mathrm{i}}$ is found by assuming that the narrowest line width is from an ion species with $T_{\mathrm{i}}=0$. This determines the maximum $v_{\mathrm{nt}}$ allowing the lower bound $T_{\mathrm{i}}$ for all the other ions to be inferred. 
	
\begin{figure}
\centering \includegraphics[width=0.9\textwidth]{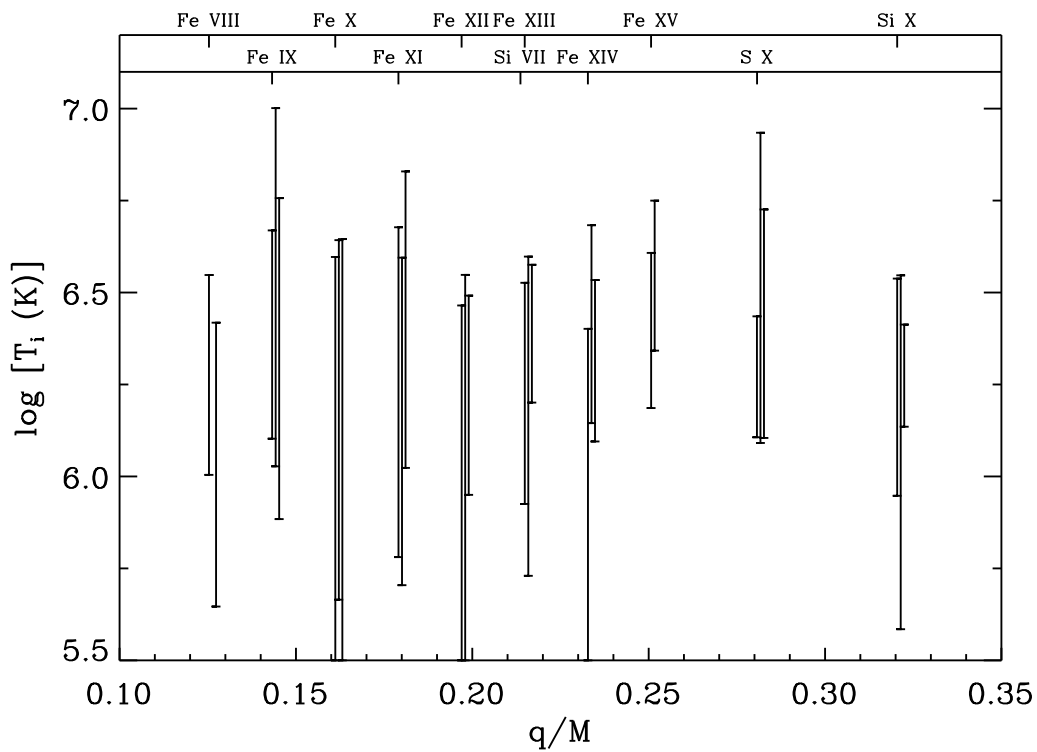}
\caption{\label{fig:ti} Upper and lower bounds for $T_{\mathrm{i}}$ as a function of charge-to-mass ratio $q/M$ for several spatial bins of the 21-Oct-2007 observation. Data from three representative spatial bins are slightly staggered along the x-axis for clarity. The coordinates of the selected bins are, from left to right, $(1069^{\prime\prime}, -158^{\prime\prime})$, $(1069^{\prime\prime},-238^{\prime\prime})$, and $(1039^{\prime\prime}, -238^{\prime\prime})$. These measurements indicate that all of the ions have about the same temperature in the quiet Sun.
}
\end{figure}

	Figure~\ref{fig:ti} shows the inferred bounds for $T_{\mathrm{i}}$ for various ion species at three representative isothermal locations in the equatorial observation. The $T_{\mathrm{i}}$ results are plotted versus the charge-to-mass ratio $q/M$. The reason for this is that in coronal holes $T_{\mathrm{i}}$ has been found to vary with $q/M$ \citep{Landi:ApJ:2009, Hahn:ApJ:2010}. On the other hand, these quiet Sun data are consistent with all of the ion species having the same $T_{\mathrm{i}}$ in the range of $\approx 1$--$3$~MK. 
	
\begin{figure}
\centering \includegraphics[width=0.9\textwidth]{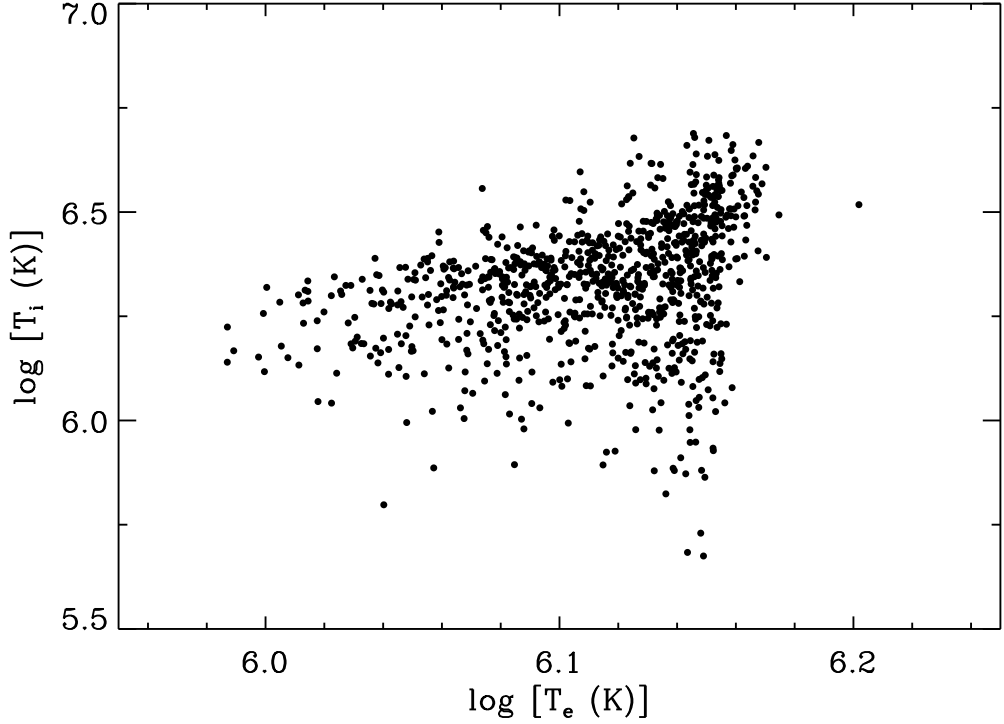}
\caption{\label{fig:ticorr} $T_{\mathrm{i}}$ versus $T_{\mathrm{e}}$ in the equatorial observation. Here $T_{\mathrm{i}}$ was found by assuming that at a given location all the ion species have the same $T_{\mathrm{i}}$ and $v_{\mathrm{nt}}$. This correlation was performed for each bin where the DEM indicated the plasma was nearly isothermal. The results show that there is a correlation between the ion and electron temperatures (see text). The uncertainty on the individual $T_{\mathrm{i}}$ measurements is typically 0.1--0.4 in the dex, depending on the height in the observation. It is large because there are only a few ions having different masses. The uncertainty for the electron temperature is taken to be $\sigma_{T}$ from the DEM analysis and is $< 0.1$ in the dex. 
}
\end{figure}

	It is possible to estimate $T_{\mathrm{i}}$ directly, if we assume that at a given location all of the ions have the same $T_{\mathrm{i}}$ and the same $v_{\mathrm{nt}}$. Since in each spatial bin there are ions from several different elements, having different masses, we can perform a least squares fit to Equation~(\ref{eq:veffdefine}) to find the parameters $T_{\mathrm{i}}$ and $v_{\mathrm{nt}}$ throughout the observation. This analysis yields an average $T_{\mathrm{i}}$ in the equatorial observation of $T_{\mathrm{i}} \approx 2.1 \pm 1.0$~MK and in the high latitude observation $T_{\mathrm{i}} \approx 1.8 \pm 0.8$~MK. To within the uncertainties, these are similar to the inferred electron temperatures of $T_{\mathrm{e}} \approx 1.1$--$1.8$~MK . The uncertainties in $T_{\mathrm{i}}$ are large because there are only a few different masses among the measured ions and only a few lines are bright enough to have reliable line width measurements up to large heights. However, we did find a clear correlation between $T_{\mathrm{i}}$ and $T_{\mathrm{e}}$ measured in each spatial bin of our data where the DEM indicated the region was isothermal (Figure~\ref{fig:ticorr}). The strength of the correlation can be quantified using the non-parametric Spearman rank-order correlation coefficient $\rho_{c}$, which takes values between -1 and 1. Here we find a positive correlation, $\rho_{c} = 0.24$ with a significance of greater than 99\%, meaning that the probability that this correlation arose by chance is very unlikely. Based on all the above results, we assume in the analysis of $v_{\mathrm{nt}}$ that $T_{\mathrm{i}}=T_{\mathrm{e}}$, but also consider as a systematic error that $T_{\mathrm{i}}$ maybe slightly larger.	
	
\subsection{Magnetic Field}\label{subsec:mag}
	
	The magnetic field in the corona was determined using a potential field source surface (PFSS) model \citep{Schatten:SolPhys:1969, Wang:ApJ:1992, Schrijver:SolPhys:2003}. Such models assume that there are no currents between $r=R_{\sun}$ and the source surface at $r=R_{s}$, where $R_{s}=2.5$~$R_{\sun}$ has usually been found to produce good agreement with observations \citep{Hoeksema:JGR:1983}. Between $R_{\sun}$ and $R_{s}$ the magnetic field $\mathrm{B}$ is calculated by solving the Laplace equation subject to boundary conditions set by a photospheric magnetograms at $r=R_{\sun}$ and an assumption of radial field lines at $R_{s}$. Despite these simplifications, PFSS models have been shown to accurately reproduce the large scale structures of the magnetic field outside of active regions \citep{Wang:ApJ:1992, Neugebauer:JGR:1998, Schrijver:SolPhys:2003, Riley:ApJ:2006}. 	
	
	We have used the PFSS package available in \textit{solarsoft} and described by \citet{Schrijver:SolPhys:2003}. The data are available at six hour intervals and we chose the magnetic field model closest in time to our observations, 2008 Jan 21 12:04:00 for the equatorial observation and 2007 Oct 21 18:04:00 for the high latitude observation. We also note that we chose our observations from the solar West limb, so that the magnetic field data are based on the most recent photospheric magnetograms. Observations at the East limb would be observing structures rotating around from the far side of the Sun and so rely more heavily on the flux transport model used to extrapolate the magnetograms. 
	
\subsection{Tracing Properties Along the Field}\label{subsec:trace}

	Our objective is to find the variation along the magnetic field of various properties derived from the EIS spectra. In order to do this, we first traced the field lines that pass through each spatial bin of our EIS observations. Next, we introduced several selection criteria to control for line of sight effects. Finally, we performed the analysis studying the variation of properties along the field lines meeting our criteria. 
	
	The first step then, was to trace the field lines most relevant to the EIS observations. Each spatial bin has a coordinate $(x,y)$, which is measured in arcseconds relative to the center of the Sun. The coordinate along the line of sight is $z$. Since the measured intensity is $\propto n_{\mathrm{e}}^2$ and the density decreases with height, we expect the spectrum to be dominated by emission at the point closest to the Sun along the line of sight. Therefore, we traced the field lines passing through the coordinates $(x,y,0)$, which is the plane passing through the center of the Sun. The field lines, though, generally do not remain precisely in this plane, as discussed below. We should also note that the PFSS model uses heliographic coordinates $(r, \theta, \phi)$, which can be converted to Cartesian coordinates using the formulae given by \citet{Thompson:AA:2006}.

	Although the point $z=0$ should make the greatest contribution to the emission, each observed spectrum is a sum over all the emission along the line of sight. To control for these line-of-sight effects we have adopted a number of criteria in order to find those points that are most likely to represent the material at $z=0$. 
	
	First, the data have been selected to come from a height range that is not affected either by structures close to the limb or by instrument scattered light at large heights. The lower boundary for the selected data is set at 1.05~$R_{\sun}$, above the limb brightening, thereby avoiding contributions from the transition region, small scale loops, and spicules. The upper boundary is set at 1.25~$R_{\sun}$. One reason for this upper boundary is that the data become noisier with increasing height, due to the decreasing density (i.e., intensity) in the corona. So, observing at moderate heights ensures that the data has a reasonable signal-to-noise level. The other basis for this limit is that instrument scattered light is not important below this height.
	
	We have estimated the level of scattered light in these observations using on-disk intensities. In previous work, we found that stray light level can be estimated to be about 2\% of the disk intensity \citep{Hahn:ApJ:2012, Hahn:ApJ:2013a}. The scattered light also has little effect on the analysis whenever the stray light contributes less than about 50\% of the total measured intensity in a given bin \citep{Hahn:ApJ:2013a}. Since the equatorial observation does not include any on-disk data, we have estimated the disk intensity from a slightly later EIS observation that looked at disk center on 2008-Jan-22 05:41 UT. Comparing these on-disk intensities to those measured in the equatorial observation, we estimate that the stray light contamination at 1.25~$R_{\sun}$ is up to 20\% for Fe~\textsc{ix}, only about 5\% for Fe~\textsc{xiii}, and falls somewhere in between these values for the other lines used in the line width analysis. For the high latitude observation, we estimated the stray light level from the average intensity in those portions of the data looking onto the disk. Here we found that at 1.25~$R_{\sun}$ the scattered light was up to 40\% of the total intensity for Fe~\textsc{ix}, about 5\% for Fe~\textsc{xiii} and less than 25\% for the other lines used in the analysis. Note that the stray light fraction depends on the ion emitting the line. Lines formed at lower temperatures have a shorter scale height and so the real coronal emission falls off more sharply with height. Thus, the relative contribution of stray light increases faster for cooler lines. Below 1.25~$R_{\sun}$, the amount of stray light in both observations is expected to have negligible effect on the inferred line widths and can be neglected. 
	
	The second criterion for the analysis is that the observed spectrum comes from a plasma that is close to isothermal. If the plasma is not isothermal, it suggests that there are multiple emitting components along the line of sight having different temperatures. In our observations, examples of this are the prominences. A DEM analysis shows that these structures have a significant contribution from cooler material (Figure~\ref{fig:multithermal}). These structures are omitted from the analysis by requiring that the $\sigma_{T}$ of the DEM be smaller than a certain value. This cutoff was determined from the histogram of the $\sigma_{T}$ for each observation. For the equatorial observation, the cutoff was set to $\sigma_{T} < 0.1$ and for the high latitude observation we required $\sigma_{T} < 0.06$. Applying these cutoffs removes the prominence regions from the analysis. It is not clear why the cutoffs are different in each observation. However, since the DEM implicitly assumes an isothermal plasma, $\sigma_{T}$ for multithermal plasmas does not have a physical meaning since the fitting function is not valid in the first place. The broad $\sigma_{T}$ is an indicator that a different form for $\phi(T_{\mathrm{e}})$ should be used for the prominences. A more general DEM analysis of the prominence areas shows that $\phi(T_{\mathrm{e}})$ is nearly flat below the peak temperature. 
	
	A third criterion is that the magnetic field line should not deviate too far from the $z=0$ plane. As discussed above, the spectra are expected to be dominated by the point closest to the plane at $z=0$. Thus, if a magnetic field line has comes too far out of this plane, the observed spectra are unlikely to correspond to material on that field line. To mitigate this problem, we ignore points in the analysis where the height of the point on the field line at $(x,y,z)$ was greater than one intensity scale height $H_{I}$ above the point in the plane at $(x,y,0)$, i.e., we ignore points where $\sqrt{x^2+y^2+z^2}-\sqrt{x^2+y^2} > H_{I}$. We used a uniform intensity scale height of $H_{I} = 0.045$~$R_{\sun}$, corresponding to a temperature of $1.3$~MK. Note that the density $n_{\mathrm{e}} \propto \exp(-r/H)$, where $H$ is the density scale height. Since the intensity is proportional to $n_{\mathrm{e}}^2$, the intensity scale height is half the density scale height, i.e. $H_{I}=H/2$. 
	
	A condition is also imposed on the direction of the magnetic field with respect to the line of sight. Alfv\'enic waves are expected to propagate along the field lines, inducing a non-thermal width perpendicular to the magnetic field direction. The angle between the line of sight and the magnetic field direction is $\alpha = \cos^{-1}(\mathbf{\hat{b}} \cdot \mathbf{\hat{z}} )$, where $\mathbf{\hat{b}}$ is the unit vector in the direction of the magnetic field and $\mathbf{\hat{z}}$ is the unit vector along the line of sight. In order to observe Alfv\'enic waves, the field must be nearly perpendicular to the line of sight, that is, $\alpha$ must be close to 90$^{\circ}$. This criterion is easily satisfied by the equatorial observation, where we required that the angle between the magnetic field and the line of sight be within $20^{\circ}$ of perpendicular. However, for the high latitude observation this criteria would rejects a large fraction of the data. So we allowed the magnetic field line to vary up to $35^{\circ}$ from the line of sight and instead of using a very strict limit on $\alpha$ for the high latitude observation, we have checked the influence of $\alpha$ on the analysis by correcting for the angle, as described below. 
	
	Requiring the observed region to be perpendicular to the line of sight also reduces the possible line broadening due to flows. Flows could be driven along the field by cooling plasma falling down from the corona, heated plasma rising into the corona, or a siphon flow caused by pressure differences between the two footpoints of the loop. Such flows would be observed as a Doppler shift when looking along a single loop, but for a collection of loops with random flows the effect would be to increase the line width. Since the flows are expected to be along the magnetic field, we reduce this possible additional line broadening by observing nearly perpendicular to the magnetic field so that the line broadening is dominated by transverse waves. 
	
	Finally, since all of these conditions remove a number of bins from our analysis, we require that there be enough points meeting all the criteria along each field line to reveal a meaningful trend. To this end, there must be at least five bins meeting all the criteria along each analyzed field line and these data must span a distance of at least 70~Mm (0.1~$R_{\sun}$). Note that due to the field of view of the observation, the maximum length that could be observed along any field line is $\sim 500$~Mm. 
	
	In many respects, the equatorial observation is superior to the high latitude observation. It contains less prominence material and the field lines are more nearly perpendicular to the line of sight. For these reasons, we focus our discussion below on the results of the equatorial observation and use the high latitude observation to confirm that the results also apply to other quiet Sun regions. 

\section{Results and Discussion}\label{sec:res}

\subsection{Extracting Wave Characteristics from Line Widths}\label{subsec:damp}

	The amplitude of the Alfv\'en waves can be observed indirectly through $v_{\mathrm{nt}}$, which is proportional to the amplitude of the waves \citep{Esser:JGR:1987, Hassler:ApJ:1990, McClements:SolPhys:1991}. Low frequency Alfv\'{e}n waves carry most of the wave power into the corona \citep{Cranmer:ApJS:2005, Tomczyk:ApJ:2009}. The energy flux density $F$ of these waves is related to $v_{\mathrm{nt}}$ and the mass density $\rho$ through \citep{Moran:AA:2001}
\begin{equation}
F = \rho \avg{\delta v^{2}} V_{\mathrm{A}}, 
\label{eq:Eflux}
\end{equation}
where $V_{\mathrm{A}} = B/\sqrt{4 \pi \rho}$ is the Alfv\'{e}n speed, with $B$ being the magnetic field strength, and $\avg{\delta v^{2}} = 2v_{\mathrm{nt}}^2$ is the velocity amplitude of the waves. Substituting these definitions into Equation~(\ref{eq:Eflux}) gives the relation
\begin{equation}
F = \frac{1}{\sqrt{\pi}} \rho^{1/2} v_{\mathrm{nt}}^2 B.
\label{eq:vntrel}
\end{equation}
The mass density can be found from $n_{\mathrm{e}}$ using $\rho = 1.15m_{\mathrm{p}}n_{\mathrm{e}}$, where $m_{\mathrm{p}}$ is the proton mass and the 1.15 accounts for the estimated 8.5\% solar helium abundance relative to protons \citep{Grevesse:SSR:1998}. 

	The energy flux crossing an area $A$ is given by $FA$. \citet{Moran:AA:2001} has noted that the flux tube area increases as $B$ decreases so that the factor $BA$ is a constant. Therefore if the Alfv\'{e}n wave energy is conserved, i.e., if the waves are undamped, then $FA$ is constant and Equation~(\ref{eq:vntrel}) implies 
\begin{equation}
v_{\mathrm{nt}} \propto \rho^{-1/4}. 
\label{eq:vntrho}
\end{equation}
This expression is predicted by linear (WKB) theory, but it is also valid without such an approximation for outward propagating waves when the fluid velocity is much smaller than the Alfv\'en speed \citep{Cranmer:ApJS:2005}. Equation~(\ref{eq:vntrho}) can be used to compare the behavior of $v_{\mathrm{nt}}$ with $n_{\mathrm{e}}$ measurements. Deviations of $v_{\mathrm{nt}}$ from the predicted $\rho^{-1/4}$ dependence are a signature of damped Alfv\'{e}n waves \citep[e.g.,][]{Hahn:ApJ:2013a}. 

	In order to measure the energy flux of Alfv\'enic waves in our observations and determine whether they are damped or not, we have measured $FA$. To derive $v_{\mathrm{nt}}$, we used the result that $T_{\mathrm{i}}\approx T_{\mathrm{e}}$ and subtracted the corresponding thermal velocity, $2k_{\mathrm{B}}T_{\mathrm{e}}/M$, from the measured line widths. Since $BA$ is constant, we can infer relative changes in the $A$ using the magnetic field strengths from the PFSS model. That is, 
\begin{equation}
\frac{A(s)}{A(0)}= \frac{B(0)}{B(s)}, 
\label{eq:areafac}
\end{equation}
where $s$ is the coordinate along a field line of length $2L$; $s=0$  and $2L$ at the footpoints of the loop where $r=R_{\sun}$. 
	
	As discussed above, the density was measured using line ratios. However, since this diagnostic depends on the ratio of two line intensities, it is more sensitive to noise than the intensities themselves and can only be used reliably up to $\approx 1.2$~$R_{\sun}$ in these observations. Instead, we use the EM derived from the DEM analysis. EM is proportional to $n_{\mathrm{e}}^2$ and so it can be used to measure relative changes in $n_{\mathrm{e}}$. Since the EM is derived from the DEM analysis, which aggregates more data, it can be used reliably over the full height range. An absolute magnitude for $n_{\mathrm{e}}$ was derived from the EM measurements by calibrating to the Fe~\textsc{xii} line ratio diagnostic densities at low heights, up to about 1.2~$R_{\sun}$. For this calibration, we determined the effective path length $l$ such that $\mathrm{EM}=n_{\mathrm{e}}^2 l$. 
As a further check that using the EM does not bias the results, we also performed the analysis in a more limited height range $< 1.2$~$R_{\sun}$ using $n_{\mathrm{e}}$ derived from the line ratios directly and found essentially the same results as deriving $n_{\mathrm{e}}$ from the emission measure. 
	
	We calculated $F(s)A(s)/A(0)$ along the field lines for every point in the observations meeting our selection criteria. This was done using Equations~(\ref{eq:vntrel}) and (\ref{eq:areafac}) along with $v_{\mathrm{nt}}(s)$, $n_{\mathrm{e}}(s)$, and $B(s)$ derived from the spectra and the PFSS model. For each field line, we fit these results using a function that is a constant $F_{0}$ up to some point $s_{d}$ where the damping begins, above $s_{d}$ there is an exponential decay with a damping length scale $L_{d}$: 
\begin{equation}
F A(s)/A(0) = \left\{ \begin{array}{l}
															F_{0} \mbox{ for } s < s_{d} \\
															F_{0} \mathrm{e}^{-(s-s_{d})/L_{d}} \mbox{ for } s \geq s_{d}
															\end{array} \right. .
\label{eq:fitexp}
\end{equation}
Since $F A(s)/A(0) = F_{0}$ at $s=0$, or equivalently at $r=R_{\sun}$, the quantity $F_{0}$ represents the energy flux injected into one footpoint at the base of the corona. 

\begin{figure}
\centering \includegraphics[width=0.9\textwidth]{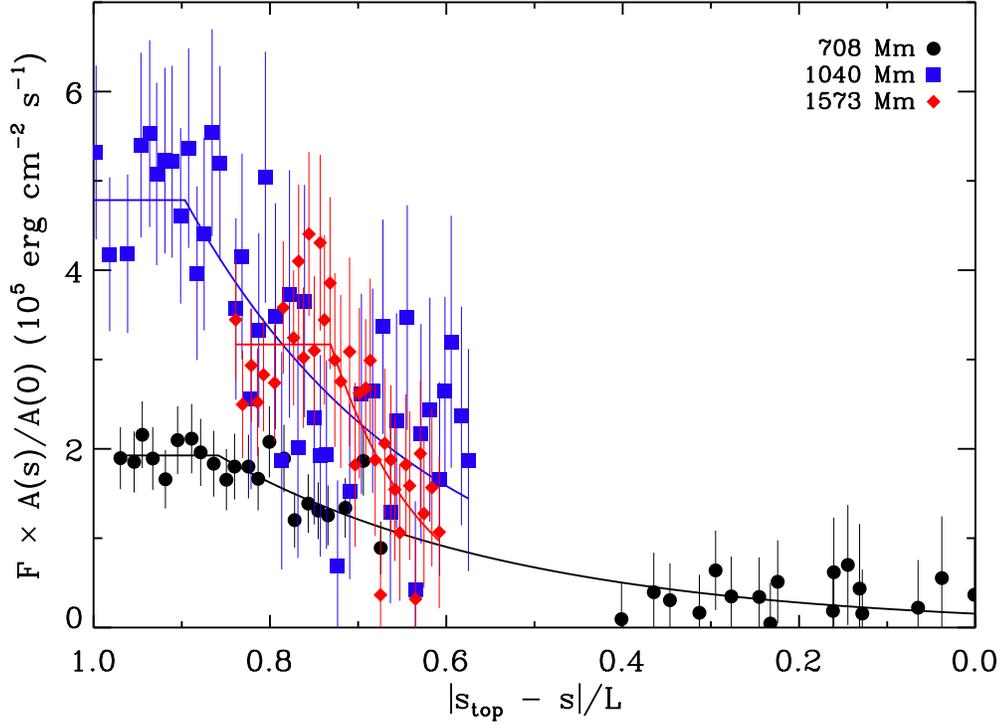}
\caption{\label{fig:fafit} Three examples are shown of the derived $FA(s)/A(0)$. These data are traced along the three field lines illustrated by the dashed lines in Figure~\ref{fig:region41}. The $x$ axis here is the distance from the top of the loop normalized by the half-length of the loop $L$. The value of $L$ is given in the caption for the different examples. The data on this plot are derived from the Fe~\textsc{xii}~195.12~\AA\ line widths.
}
\end{figure}

	Figure~\ref{fig:fafit} shows several examples of these fits, for the equatorial observation in the case where $v_{\mathrm{nt}}$ was derived from Fe~\textsc{xii}. The fits for the other measured spectral lines are essentially the same. In this plot the $x$ axis is the distance along the field line measured from the top of the loop, $s_{\mathrm{top}}$, normalized by the loop half-length $L$. The different colors indicate the fits for loops of different $L$. The loops are not necessarily symmetric and so the position $s=L$ does not generally correspond to the top of the loop. However, we normalize our results to $L$ in order to facilitate comparisons with theory, for which $L$ is usually a parameter.

	A similar analysis was performed for the high latitude observation. However, as mentioned above, the field lines traced in this observation are not quite as perpendicular to the line of sight as those in the equatorial observation. Therefore, the analysis was carried out in two ways. First, we performed the analysis exactly as for the equatorial observation, but keeping data where $\alpha$ was up to $35^{\circ}$ different from perpendicular. The second method was to correct for the rather large $\alpha$ in the observation. For this we assume that all of the waves are transverse, so that the non-thermal velocity along the field was zero. In this case, the parallel line broadening is due only to the thermal velocity $v_{\mathrm{th}}$. So the observed $v_{\mathrm{eff}}$ is given by \citep{Hahn:ApJ:2013}
\begin{equation}
v_{\mathrm{eff}}^2 = v_{\mathrm{th}}^{2} \cos^{2}(\alpha) + v_{\mathrm{eff},\perp}^2 \sin^{2}(\alpha), 
\label{eq:fitani}
\end{equation}
where $v_{\mathrm{eff},\perp}$ is the line width we would measure if the field line were perpendicular to the line of sight. It is this quantity $v_{\mathrm{eff},\perp}$ that is desired, so we corrected the original measured $v_{\mathrm{eff}}$ using
\begin{equation}
v_{\mathrm{eff},\perp} = \left[ \frac{v_{\mathrm{eff}}^2 - v_{\mathrm{th}}^2 \cos^{2}(\alpha)}{\sin^{2}(\alpha)} \right]^{1/2}. 
\label{eq:veffcorrect}
\end{equation}
Both methods, either allowing a greater range of $\alpha$ or correcting for the angle, give similar results in terms of trends, but the correction produces $v_{\mathrm{eff},\perp} > v_{\mathrm{eff}}$, resulting in a correspondingly greater estimate for $v_{\mathrm{nt}}$. 

\subsection{Wave Energy}\label{subsec:erg}

\begin{figure}
\centering \includegraphics[width=0.9\textwidth]{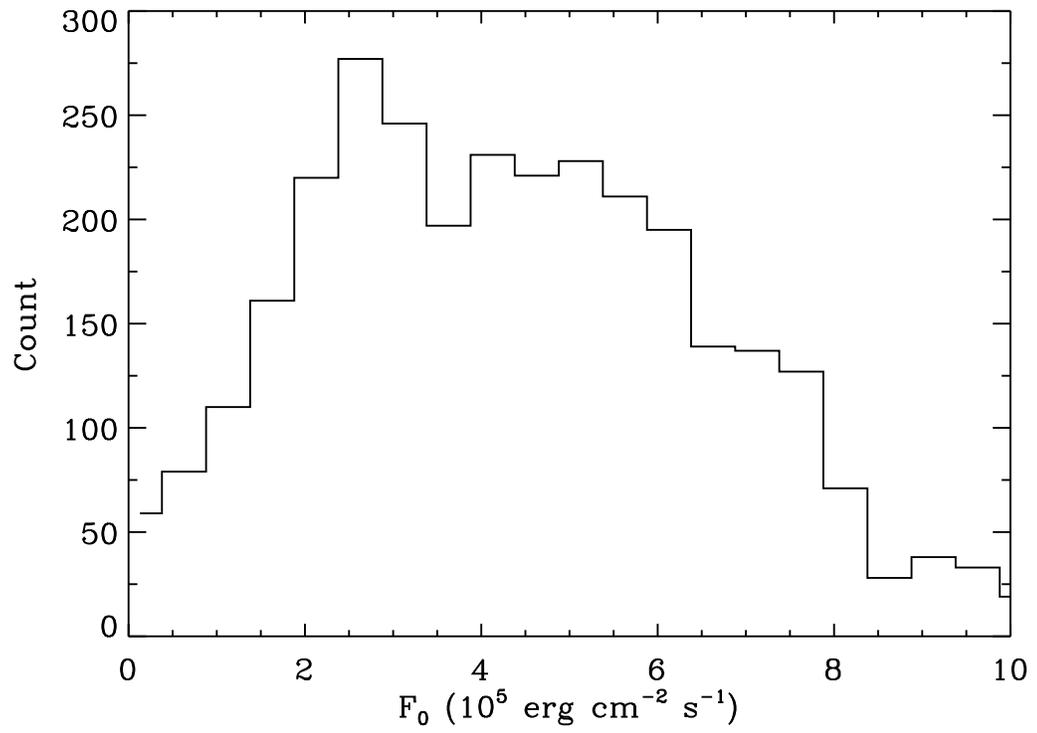}
\caption{\label{fig:fahist41} Histogram of the inferred $F_{0}$ at the base of the loops for the equatorial observation, based on aggregating all the data from the Fe~\textsc{ix}--\textsc{xiii} and Si~\textsc{x} line widths. 
}
\end{figure}

\begin{figure}
\centering \includegraphics[width=0.9\textwidth]{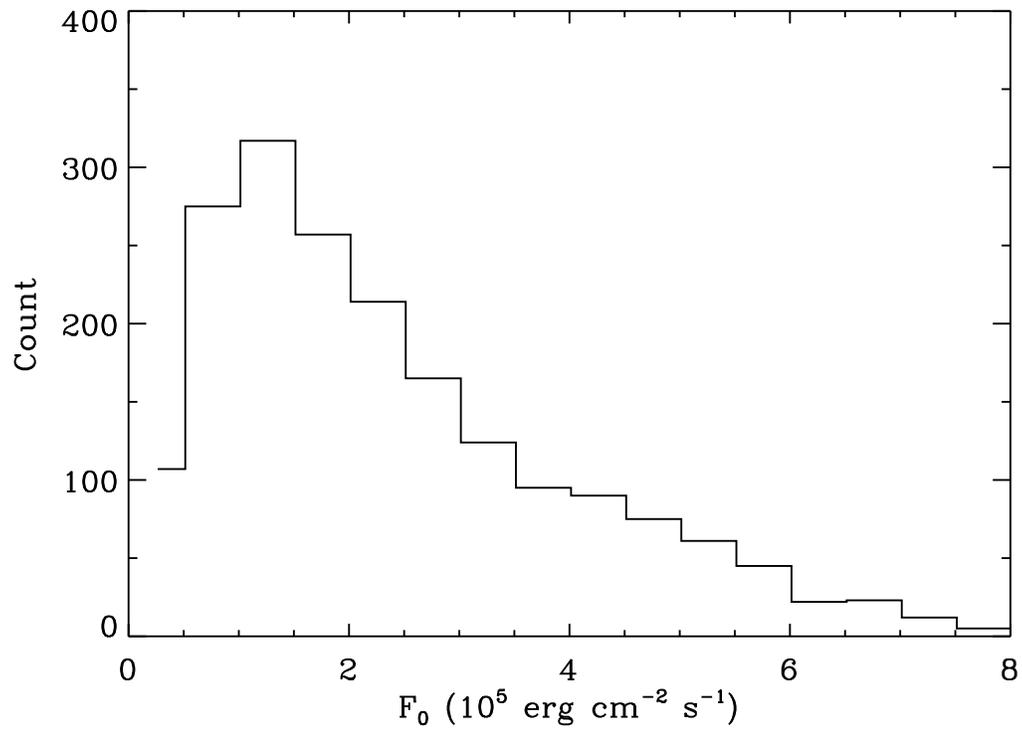}
\caption{\label{fig:fahist47} Same as Figure~\ref{fig:fahist47}, but for the high latitude observation. These results were obtained after correcting for the angle between the line of sight and the magnetic field.
}
\end{figure}

\begin{figure}
\centering \includegraphics[width=0.9\textwidth]{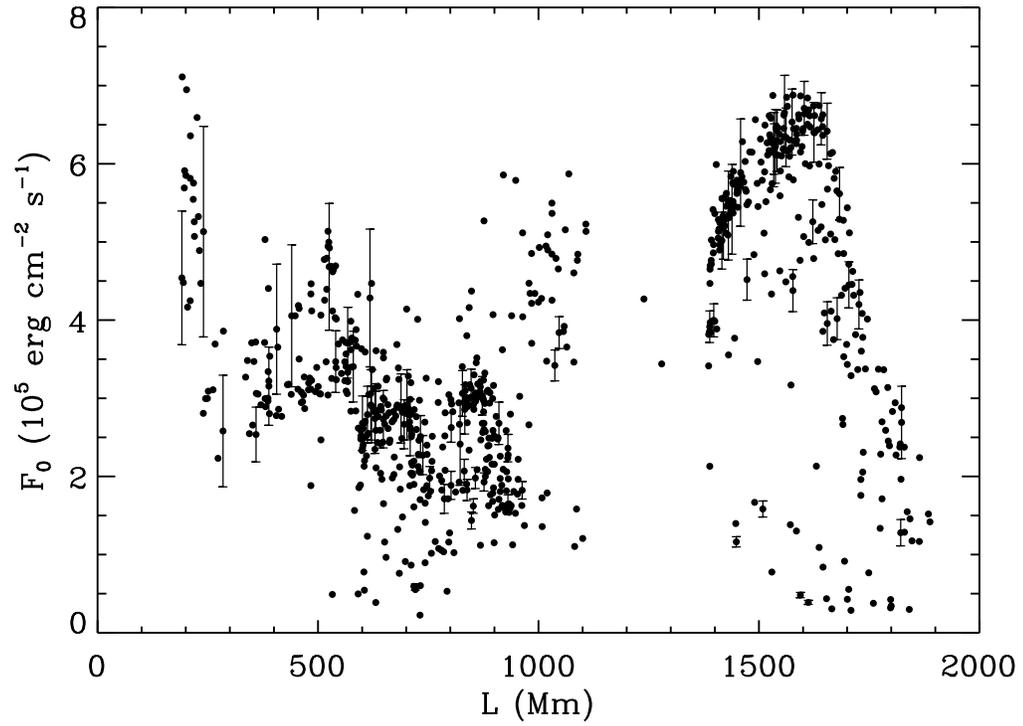}
\caption{\label{fig:fvsl} The base energy flux $F_{0}$ versus the loop half length $L$. These data are from the analysis of the Fe~\textsc{xii}~195.12~\AA\ line widths for the equatorial observation. Error bars are plotted only for a few selected points.
}
\end{figure}

	For the equatorial observation we find that on average $F_{0} = 5.2 \pm 1.3 \times 10^{5}$~$\mathrm{erg\,cm^{-2}\,s^{-1}}$. The results were similar for the northwest observation with $F_{0} = 3.4 \pm 0.8 \times 10^{5}$~$\mathrm{erg\,cm^{-2}\,s^{-1}}$ with no correction for $\alpha$ and $F_{0} = 4.8 \pm 1.3 \times 10^{5}$~$\mathrm{erg\,cm^{-2}\,s^{-1}}$ with the correction. These averages are taken over all six spectral lines from which we analyzed line widths, all of which gave similar results. Histograms of the $F_{0}$ results are shown in Figures~\ref{fig:fahist41} and \ref{fig:fahist47} for the equatorial and for the high latitude observation with the $\alpha$ correction, respectively. We did not find any clear dependence of $F_{0}$ on the loop length, which implies that the energy input into the corona is roughly similar everywhere. This is illustrated in Figure~\ref{fig:fvsl}, which shows the inferred $F_{0}$ versus the loop half-length $L$ from the analysis of the Fe~\textsc{xii} data.
	
	There are also several sources of systematic error to consider. First, in finding $F$ we used densities from Fe~\textsc{xii}. The Fe~\textsc{xiii} density diagnostic infers densities that are a factor of two smaller. If those values are more accurate then our $F_{0}$ should be reduced by a factor of $\sqrt{2}$. Additionally, we have assumed that $T_{\mathrm{i}}= T_{\mathrm{e}}$ when inferring $v_{\mathrm{nt}}$ from $v_{\mathrm{eff}}$. However, our $T_{\mathrm{i}}$ measurements are also consistent with $T_{\mathrm{i}}$ being slightly larger than $T_{\mathrm{e}}$. To account for the uncerainty in $T_{\mathrm{i}}$, we performed the analysis also for $T_{\mathrm{i}}=2T_{\mathrm{e}}$. This reduces the inferred $F_{0}$ by about a factor of two. Thus, in total these systematic uncertainties imply that the energy flux at the base of the corona may be up to a factor of $2.8$ smaller than the results given above. For the equatorial observation this corresponds to  $F_{0}=1.8 \times 10^{5}$~$\mathrm{erg\,cm^{-2}\,s^{-1}}$ and similarly for the corrected high latitude observation $F_{0} = 1.7 \times 10^{5}$~$\mathrm{erg\,cm^{-2}\,s^{-1}}$. 
	
	One further issue is that we have not actually determined the wave mode that causes the non-thermal velocity. Since we observe line widths perpendicular to the magnetic field, we see transverse Alfv\'enic waves. Such waves may be may be torsional Alfv\'en waves or swaying fast kink mode waves. Our line width observations cannot distinguish between these two types of waves, but instead see the total broadening due to both types of waves. This contrasts with Doppler shift measurements, such as those of \citet{Tomczyk:ApJ:2009} that are sensitive to kink mode oscillations but not torsional Alfv\'en waves. 
	
	If kink waves contribute to our measurements, then using Equation~(\ref{eq:Eflux}) overestimates the amount of energy carried by the waves. This is because for kink waves, the energy is not distributed uniformly throughout the volume. \citet{Goossens:ApJ:2013} has shown that for kink waves travelling along a uniform density cylinder having a mass density $\rho_{\mathrm{i}}$ inside the cylinder and $\rho_{\mathrm{e}}$ external to the cylinder, Equation~\ref{eq:Eflux} overestimates the energy flux by a factor of
\begin{equation}
\eta^{2} \frac{\rho_{\mathrm{i}}}{\rho_{\mathrm{i}}+\rho_{\mathrm{e}}}, 
\label{eq:goossens}
\end{equation}
where $\eta$ is a measure of the inverse filling factor. For the corona we can estimate that $\rho_{\mathrm{i}} \approx 2\rho_{\mathrm{e}}$ to $6 \rho_{\mathrm{e}}$ \citep{November:ApJ:1996, Raymond:ApJ:2014} and $\eta \approx 2$ \citep{Raymond:ApJ:2014}. If we observed only kink waves, this would imply that we have overestimated the energy flux by a factor of about 3. \citet{DePontieu:ApJ:2012} has measured both torsional waves and kink waves in spicules and found that there is roughly equal energy in both types of waves. Assuming that the corona also has half of the energy content in torsional Alfv\'en waves, for which Equation~\ref{eq:Eflux} is reasonable, our estimate of the wave energy flux should be reduced by about a factor of about 1.5. This implies that if all of our systematic uncertainties combine to reduce the wave energy, the waves carry an energy flux of at least $1.2 \times 10^{5}$~$\mathrm{erg\,cm^{-2}\,s^{-1}}$.  
	
	It has been estimated that quiet Sun regions require an energy input of about $3\times10^{5}$~$\mathrm{erg\,cm^{-2}\,s^{-1}}$ \citep{Withbroe:ARAA:1977}. Here we have estimated the energy injected at each footpoint of a quiet Sun loop to be, on average $1.2$ -- $5.2 \times 10^{5}$~$\mathrm{erg\,cm^{-2}\,s^{-1}}$. The total energy input onto a loop is twice this value because each loop has two footpoints. Therefore, our results indicate that waves probably carry at least 80\% of the energy into the corona needed to heat quiet Sun regions. 
	
\subsection{Wave Damping}\label{subsec:damping}	
	
	We have also found that the waves are damped in these quiet Sun regions. One parameter to characterize this damping is $s_{d}$ from Equation~(\ref{eq:fitexp}), which represents the distance from the footpoint of the loop above which $v_{\mathrm{nt}}$ is no longer proportional to $\rho^{-1/4}$. This parameter can only be determined accurately when it falls within the observed region. Since our data only cover a limited height field of view, it is possible that for long or short loops, $s_{d}$ may not be observed. For example, damping may start below the range measured in the observation, meaning that $FA(s)/A(0)$ was decreasing everywhere in the measured range. Alternatively, if $s_{d}$ is above the measured range, it implies that $FA(s)/A(0)$ is constant throughout our data.
	
\begin{figure}
\centering \includegraphics[width=0.9\textwidth]{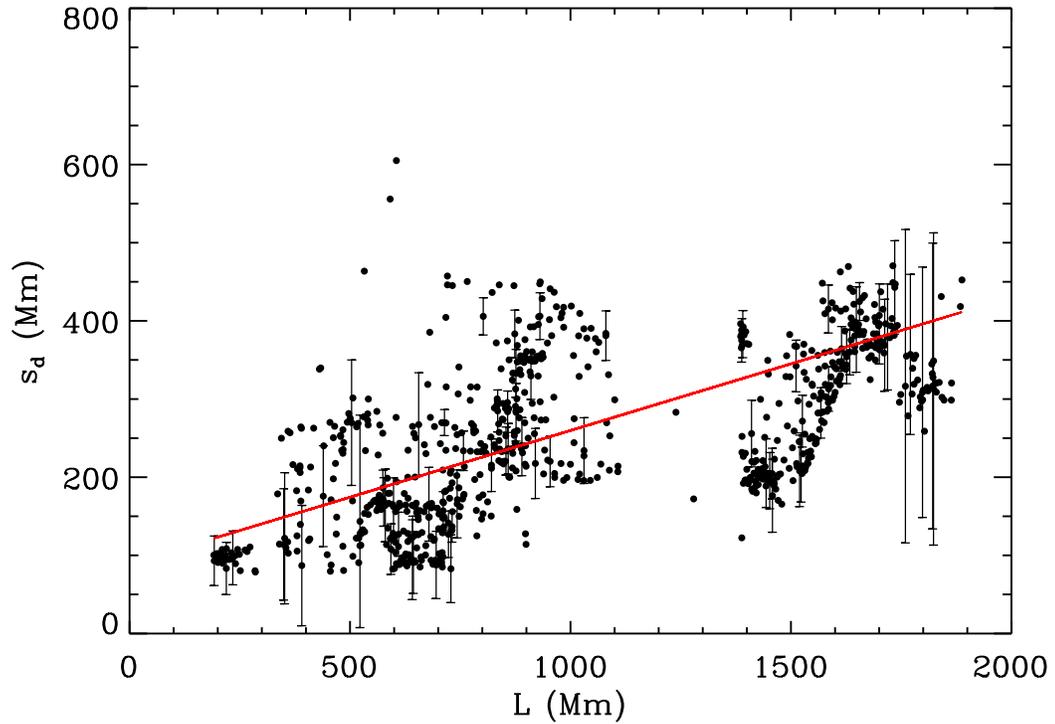}
\caption{\label{fig:sd41} The distance $s_{d}$ from the footpoint at which the damping begins versus the loop half-length $L$ for the equatorial observation. Error bars are plotted for selected points. The solid line is a linear fit to the data. The correlation indicates that damping ``turns on'' at a longer distance in larger loops. The data plotted here are derived from the line widths of the Fe~\textsc{xii}~195.12~\AA\ line.
}
\end{figure}

\begin{figure}
\centering \includegraphics[width=0.9\textwidth]{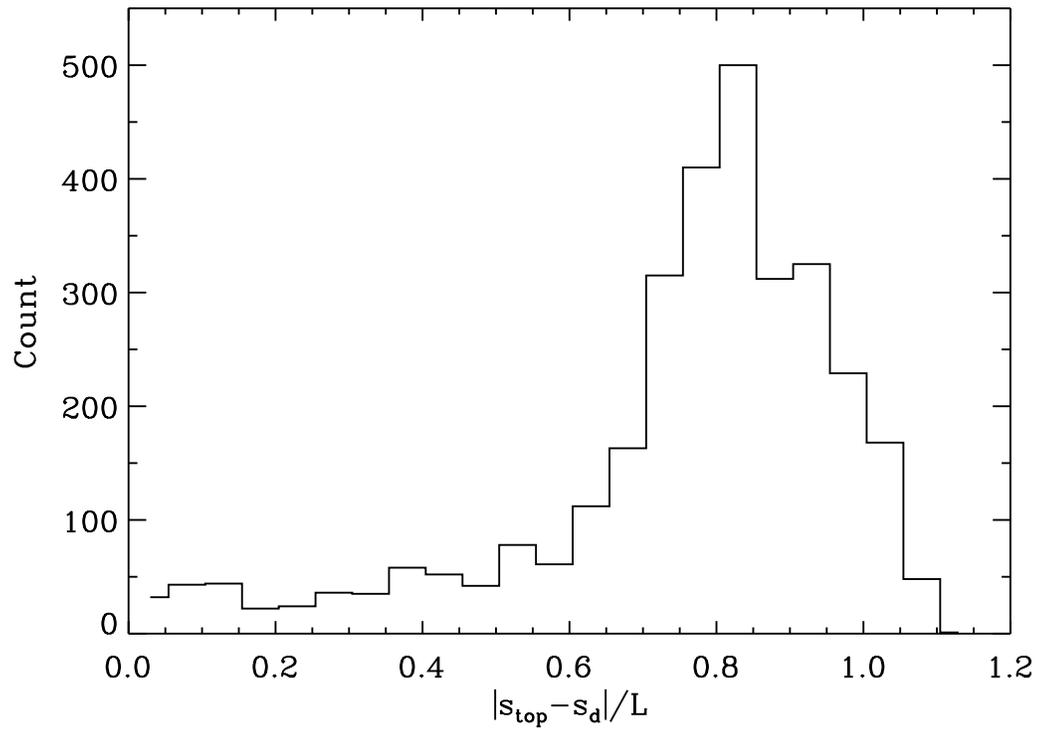}
\caption{\label{fig:sdhist41} Histogram of the fraction of the loop above the damping point, $|s_{\mathrm{top}}-s_{d}|/L$, for the equatorial observation. The histogram aggregates all the data from the various measured line widths.
}
\end{figure}

	For those field lines where $s_{d}$ fell within the field of view, we found a clear correlation between the position  at which damping begins and the size of the loop. Figure~\ref{fig:sd41} illustrates this trend for the equatorial observation and shows that there is a clear correlation between $s_{d}$ and the half-length $L$. Though Figure~\ref{fig:sd41} plots only data from the Fe~\textsc{xii} line, a similar correlation is observed consistently for every spectral line measured. Here we find a strong positive correlation, with Spearman correlation coefficient $\rho_{c} = 0.65$ at a significance of greater than 99\%. Figure~\ref{fig:sd41} suggests that the relationship is nearly linear, which implies that the waves are damped over a constant fraction of the loop length. Assuming that the loops are symmetric, the damping must be centered on the loop top and occur over a distance $2\left(s_{\mathrm{top}}-s_{d}\right)$. For the equatorial observation, the average fraction of the $2L$ loop length over which the damping occurs is $|s_{\mathrm{top}}-s_{d}|/L = 0.8 \pm 0.1$. Figure~\ref{fig:sdhist41} illustrates a histogram of this fraction, which aggregates all the spectral data from Fe~\textsc{ix}--\textsc{xiii} and Si~\textsc{x}. 
	
\begin{figure}
\centering \includegraphics[width=0.9\textwidth]{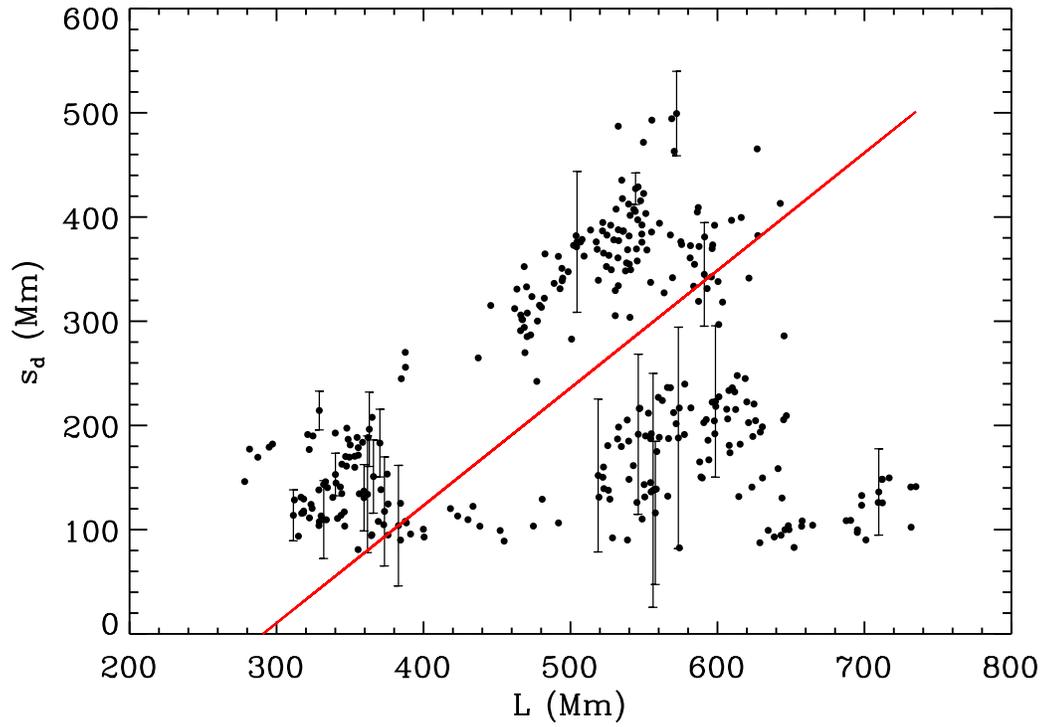}
\caption{\label{fig:sd47} Same as Figure~\ref{fig:sd41}, but for the high latitude observation. Here, there appear to be two classes of loops. All of the loops have a strong correlation between $s_{d}$ and $L$, but there is an offset between the two classes so that for some loops the damping is more concentrated near the loop top. 
}
\end{figure}

\begin{figure}
\centering \includegraphics[width=0.9\textwidth]{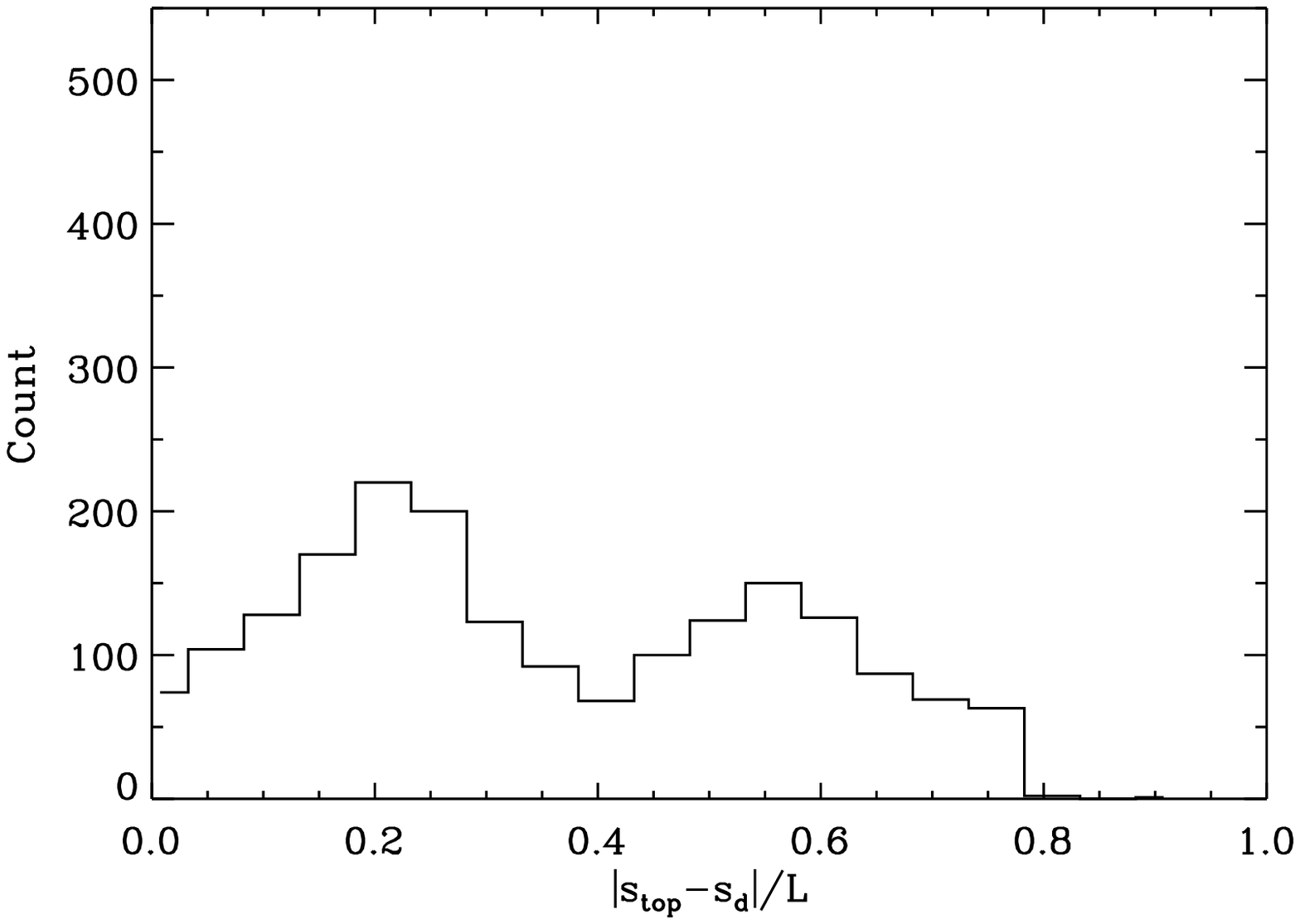}
\caption{\label{fig:sdhist47} Same as Figure~\ref{fig:sdhist41}, but for the high latitude observation. 
}
\end{figure}

	In the high latitude observation, there is a similar correlation between $s_{d}$ and $L$ (Figure~\ref{fig:sd47}). For this observation the correlation is somewhat weaker with $\rho_{c} =0.18$, but the correlation is still greater than 99\% significant. There appear to be two classes of loops, where each has a linear dependence between $s_{d}$ and $L$, but with an offset between the two groups, implying that the damping is more concentrated near the loop top in one group than in the other. This is most clearly seen in the histogram in Figure~\ref{fig:sdhist47}. There are two peaks in the distribution of $|s_{\mathrm{top}}-s_{d}|/L$, one at $\approx 0.2$ and another at $\approx 0.55$. This implies that for some loops the damping is concentrated more strongly at the loop top than for others. It is not clear why there are two groups of loops. We have looked at the geometry of the loops in terms of their length, roundness, and symmetry but have found no relation between these properties and the $|s_{\mathrm{top}}-s_{d}|/L$. There is also no correlation with the base energy flux $F_{0}$. 
	
	We have verified that the relation between $s_{d}$ and $L$ is not sensitive to our assumption that the data can be fit by Equation~(\ref{eq:fitexp}). In order to do this, we fit the line width data $v_{\mathrm{eff}}(s)$ using a sequence of two linear functions, and found the point where there was a break from a positive slope to a negative slope. For example,  Figure~\ref{fig:veffvsr} shows that $v_{\mathrm{eff}}$ is increasing at low heights, but decreases at larger heights. The position of the break has the same strong correlation with the length and height of the loop and occurs at essentially the same point as $s_{d}$ from the exponential fit. 

	 The above results are for the majority of cases where $s_{d}$ fell within the field of view of our data. We excluded from our analysis the very few cases where no damping was observed within the field of view. For the equatorial observation, there were 1408 total field lines traced, because we started with $22 \times 64$ spatial bins. Of these, 1200 met all the criteria described in Section~\ref{subsec:trace}. Only 15 of these had fits that implied that $s_{d}$ was above the measured range of $s$. This means only 1\% of the measurements were consistent with no damping observed within the field of view. These fits corresponded to field lines that were longer than average. The average $L$ for field lines with flat $FA(s)/A(0)$ in the measured range was $L=1200$~Mm compared to the average in $L$ for all the data of $L=1000$~Mm. The field of view of our observations is $512^{\prime\prime} \times 220^{\prime\prime}$, so the maximum length along a field line we can observe is around 500~Mm. Since $s_{d}\propto L$, it may be that damping on very long field lines begins at a larger $s$ than we observed. 
	 
	 There were, however, a significant number of cases where $s_{d}$ was below the minimum measured $s$. Since $s_{d}$ could not be accurately determined, these were also excluded in the analysis described above. For these cases damping may have begun closer to the footpoint of the loop than was covered by our data. Out of the 1200 total good field lines, 272 field lines had $FA(s)/A(0)$ always decreasing in the observed height range. One possibility is that these loops follow the same trend with $L$ as observed for longer ones, but that since we only observed above 1.05~$R_{\sun}$ only the top portion of the loops were seen so that damping was always present. This interpretation is consistent with the fact that for these loops the average length was $\approx 800$~Mm in the equatorial observation, which is shorter than the average length in that observation. It is also possible that the damping in these loops has a different character than in the longer loops. One possibility is base heating, where any damping of waves starts immediately above the footpoints. Such heating has been proposed to explain tomographic observations that have found quiet Sun loops with negative temperature gradients \citep{Huang:ApJ:2012}. 

\begin{figure}
\centering \includegraphics[width=0.9\textwidth]{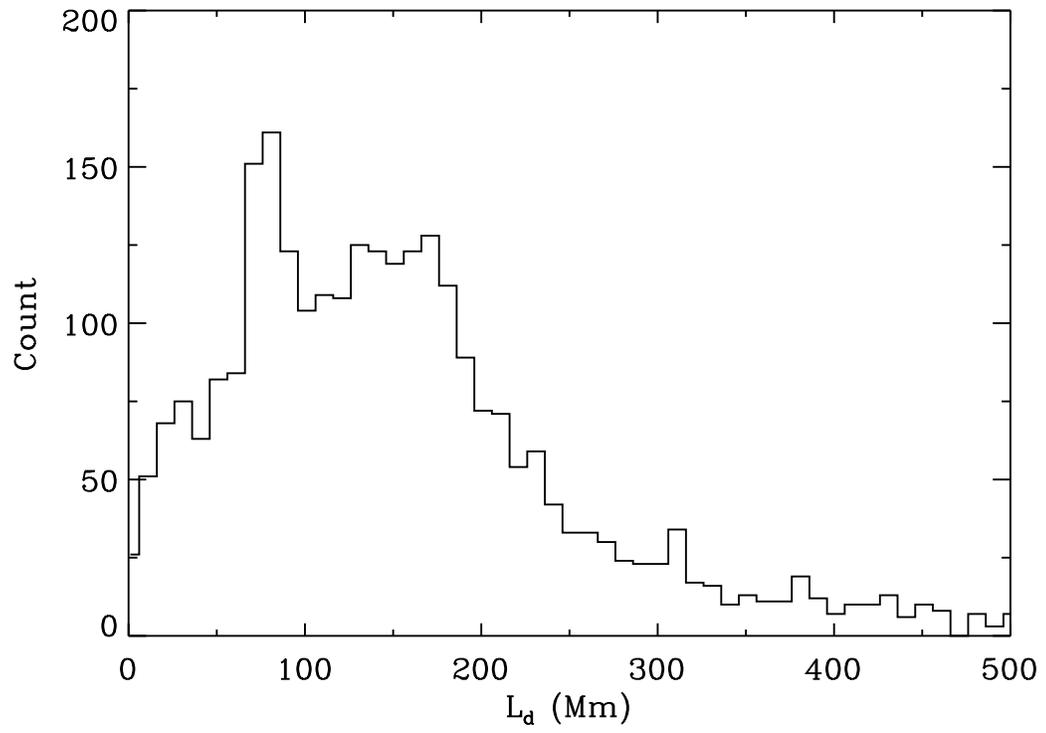}
\caption{\label{fig:ldhist41} Histogram of the damping scale length $L_{d}$ for the equatorial observation. 
}
\end{figure}

\begin{figure}
\centering \includegraphics[width=0.9\textwidth]{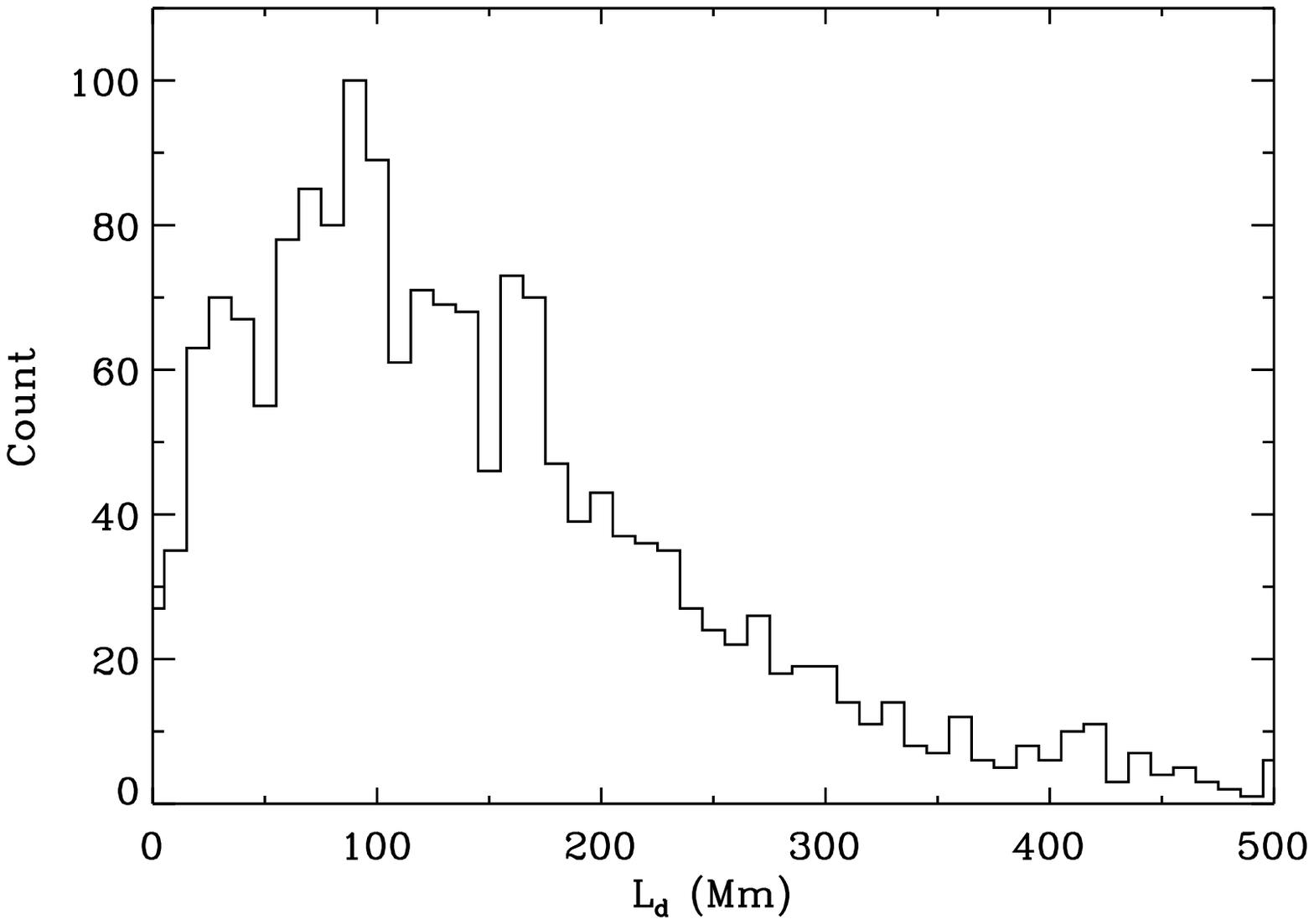}
\caption{\label{fig:ldhist47} Same as Figure~\ref{fig:ldhist41}, but for the high latitude observation.
}
\end{figure}

	Returning to the subset of 913 loops where $s_{d}$ fell within the field of view, we found the median damping length $L_{d}$ was in the range of $\sim 100$--$200$~Mm (Figures~\ref{fig:ldhist41} and \ref{fig:ldhist47}). There were no clear correlations with the loop length or height. There is a large spread in the derived $L_{d}$, which may be because the length of the loops is significantly larger than the field of view of our observations. Except for very short loops, we only see the lower fraction of the loops. This means that, although we can usually see the point where the damping begins, we do not have much data coverage above the damping point to constrain $L_{d}$. Desite this imprecision, it is clear that $L_{d}$ is shorter than $L$ except for very short loops. This implies that much of the wave energy is dissipated before the waves can reach the other footpoint. Conversely, on short loops, $L \lesssim L_{d}$, wave energy may flow from one footpoint to the other. Our findings are consistent with CoMP observations that also inferred wave damping in the quiet Sun. In those observations significant wave power was found going up along quiet Sun loops, but there was very little energy in downward propagating waves except on the shortest loops, which suggests that the waves were dissipated as they propagate from one footpoint to the other \citep{Tomczyk:ApJ:2009}.
	
	The finding that $s_{d}$ is strongly correlated with the length of the loop constrains the possible damping mechanism for the waves. For example, this finding is inconsistent with collisional damping, such as due to viscosity and resistivity. In those cases, the waves would gradually decay as they propagate, starting from the loop footpoints. Instead, damping seems to depend on global properties of the entire loop. There are several possible processes that may be able to explain our observations. 
	
	One damping mechanism that could be consistent with our results is dissipation by turbulence. Counterpropagating Alfv\'en waves can generate such turbulence. In closed coronal loops, this is likely to occur near the loop top due to the interaction of the waves launched in the opposite direction from each footpoint. This scenario has formed the basis for a unified treatement of coronal heating in both open and closed regions, in which the heating of the corona in closed magnetic fields is due to this turbulence from waves launched at the footpoints \citep{Oran:ApJ:2013, VanDerHolst:ApJ:2014}. \citet{DeMoortel:ApJL:2014} have also found evidence for Alfv\'enic turbulence in closed loops using CoMP. They found that the power spectrum of fluctuations at the top of quiet Sun loops was broader than at the footpoints, implying a turbulent cascade of wave energy to higher frequencies. 

	The damping could also occur through the resonant absorption of the kink waves. \citet{Verth:ApJ:2010} have shown that this mechanism can account for the damping seen in the CoMP results. Resonant dissipation occurs because kink waves propagate on flux tubes where the density varies with radius away from the axis of the flux tube. At some layer between the axis and the outer radius of the flux tube, there is a resonance where the kink wave frequency is equal to the local Alfv\'en frequency. This allows the kink wave to excite torsional Alfv\'en waves on that surface. For kink waves with various frequencies, there will be many such surfaces, which can interact by phase mixing and cause a cascade of wave energy to small scales where the energy is dissipated. Our measured damping lengths of $100$--$200$~Mm are consistent with what is expected from resonant damping. Although, it is not clear whether resonant damping can explain why $s_{\mathrm{d}}$ appears to be correlated with the loop length. 
	
\section{Conclusion}\label{sec:con}

	We have used a magnetic field model to aid in interpreting spectroscopic line width measurements in the quiet Sun. From the model, we are able to identify the direction of the magnetic field lines and thereby trace $v_{\mathrm{nt}}$ along the fields. Line-of-sight ambiguities were controlled for, as far as possible, by selecting data from isothermal regions where the field lines remained close to perpendicular to the line of sight. 

	Based on these data we inferred a wave energy flux starting from the base of quiet Sun field lines of about $5.2 \times 10^{5}$~$\mathrm{erg\,cm^{-2}\,s^{-1}}$. Even if the various systematic uncertainties that we discussed combine to reduce the wave energy, then we still find  $1.2 \times 10^{5}$~$\mathrm{erg\,cm^{-2}\,s^{-1}}$. This energy is present at each footpoint of the quiet Sun loops. The estimated amount of energy required to heat the quiet Sun is $3\times10^{5}$~$\mathrm{erg\,cm^{-2}\,s^{-1}}$. Thus, our findings indicate that the combined energy in Alfv\'enic waves transmitted to the corona through both footpoints is at least 80\% of that needed to heat the quiet Sun.
	
	We have found that $v_{\mathrm{nt}}(s)$ at the base of the loops varies as expected for undamped Alfv\'en waves, but at a certain height the waves begin to be damped. The height and distance along the loop at which the damping begins is strongly correlated with the length of the loop, implying that the damping is controlled by properties of the entire loop rather than simply collisional dissipation. One possibility is that the damping is caused by turbulence driven by the interaction of the counterpropagating waves launched at the opposing footpoints. 
		
\acknowledgements
We thank Dr.\ Hirohisa Hara for useful discussions. MH and DWS were supported in part by the NASA Solar Heliospheric Physics program grant NNX09AB25G and the NSF Division of Atmospheric and Geospace Sciences SHINE program
grant AGS-1060194. \textit{Hinode} is a Japanese mission developed and launched by ISAS/JAXA, with NAOJ as domestic partner and NASA and STFC (UK) as international partners. It is operated by these agencies in co-operation with ESA and NSC (Norway).
 	

\clearpage

\clearpage

\clearpage

\clearpage

\clearpage

\clearpage

\clearpage

\clearpage

\clearpage

\clearpage

\clearpage

\clearpage

\clearpage

\clearpage

\clearpage

\clearpage

\clearpage

\clearpage
\bibliography{Damping}

\end{document}